\documentclass[twocolumn]{jpsj2}

\title{
The Kondo Lattice Model in Infinite Dimensions II.  \\
Static Susceptibilities and Phase Diagram
}

\author{%
Junya \textsc{Otsuki}\thanks{E-mail address: otsuki@cmpt.phys.tohoku.ac.jp}, 
Hiroaki \textsc{Kusunose}$^1$ and Yoshio \textsc{Kuramoto}
}

\inst{%
Department of Physics, Tohoku University, Sendai 980-8578\\
$^1$Department of Physics, Ehime University, Matsuyama 790-8577\\
}

\recdate{\today}

\abst{%
Magnetic and charge susceptibilities in the Kondo lattice are derived by the continuous-time quantum Monte Carlo (CT-QMC)
method combined with the dynamical mean-field theory.
For a weak exchange coupling $J$ and near half filling of the conduction band, antiferromagnetic transition occurs as signalled by 
divergence of the staggered magnetic susceptibility with lowering temperature.  With increasing $J$,
the Kondo effect suppresses the divergence, and the critical value of $J$ agrees well with Doniach's estimate which considers the RKKY interaction as competing with the Kondo effect.
For low density of conduction electrons, a ferromagnetic ordering is observed where Doniach's estimate does not work.
Around quarter filling, a charge-density-wave (CDW) transition is found.
The CDW is interpreted from the strong-coupling limit
in terms of effective repulsion between Kondo singlets. 
}

\kword{%
Kondo lattice model, charge-density wave (CDW), Doniach's phase diagram, continuous-time quantum Monte Carlo method (CT-QMC), dynamical mean-field theory (DMFT), hypercubic lattice
}

\begin{document}
\maketitle

\section{Introduction}

Due to the strong local correlations, $4f$ electrons of rare-earth ions exhibit itinerant as well as localized properties.
A single magnetic impurity in a metal gives rise to the Kondo effect, and the RKKY interaction works between a couple of moments.
Consequently, rare-earth materials, considered as systems with periodic array of the localized moments, show a variety of phases.
The 
simplest model to describe the above situation is the Kondo lattice model given by
\begin{align}
	H_{\rm KL} = \sum_{\mib{k}\sigma} \epsilon_{\mib{k}} c_{\mib{k}\sigma}^{\dag} c_{\mib{k}\sigma} 
	+ J \sum_i \mib{S}_i \cdot \mib{\sigma}^{{\rm c}}_i,
\label{eq:H_KL}
\end{align}
where $\mib{S}_i$ denotes the localized spin at $i$ site, and $\mib{\sigma}_i^{\rm c}$ is defined by $\mib{\sigma}_i^{\rm c}=\sum_{\sigma \sigma'} c_{i\sigma}^{\dag} \mib{\sigma}_{\sigma \sigma'} c_{i\sigma'}$.
With the spectrum $\epsilon_{\mib{k}}$ fixed,
the system described in eq.~(\ref{eq:H_KL}) is specified by 
two parameters: a number of conduction electrons per site, $n_{\rm c}$, and the coupling constant $J$. 

As for the one-dimensional system, the ground-state properties of the Kondo lattice have been revealed based on the diagonalization and density matrix renormalization group methods\cite{Tsunetsugu-review, Shibata-Ueda}.
Although strong quantum fluctuations suppress an antiferromagnetism in one dimension, 
a ferromagnetism can survive since the total spin 
commutes 
with the Hamiltonian.
Actually, the ferromagnetic phase is stabilized in 
the low-carrier and strong-coupling regime, while the other region is paramagnetic. 

Although an overall picture has been clarified in one dimension, these results cannot 
remain valid in higher dimensions, where antiferromagnetism takes place.
Variational methods have 
been used for a ground-state phase diagram that includes ferromagnetic and antiferromagnetic phases at weak coupling\cite{Lacroix-Cyrot, Fazekas-Muller-Hartmann}.
More reliable results 
have been obtained in two dimensions with use of a quantum Monte Carlo simulation. 
However, the simulation is restricted to the half-filled case of the conduction band because of the notorious minus sign problem\cite{Assaad}.
On the other hand, higher dimensions can be treated by means of the dynamical mean-field theory (DMFT). 
With use of the numerical renormalization-group method, the single-particle spectrum\cite{Costi-Manini} and the magnetic phase\cite{Peters-Pruschke} have been derived. 
A fundamental problem is how the Kondo effect works in the presence of many spins, which number is comparable to or even larger than that of conduction electrons\cite{nozieres}.

In a previous paper\cite{KLM1},
we have derived formulae to evaluate spatial dependences of two-particle correlations in the framework of the DMFT.
In the DMFT, the strong local correlations are rigorously taken into account\cite{Georges}. 
Therefore, we can discuss the magnetic and charge instabilities against formation of the coherent quasi-particles.
We employ the recently developed 
CT-QMC 
algorithm to solve the effective impurity Kondo model\cite{Rubtsov, Werner, Otsuki-CTQMC}.
With these framework, numerically reliable solutions can be obtained. 
We will demonstrate existence of antiferromagnetic and ferromagnetic phases as well as the CDW phase, each of
which is stabilized depending on the conduction-electron density $n_{\rm c}$.

This paper is organized as follows.
In the next section, we review the properties of the tight-binding band in the infinite-dimensional hypercubic lattice.
Numerical results of the susceptibilities are given in \S3--6 with varying the occupation number of the conduction band.
From these results, a ground-state phase diagram is constructed in \S7.
Discussions will be finally given in \S8.



\section{Infinite-Dimensional Hypercubic Lattice}

The cubic lattice can be extended to higher dimensions, which is referred to as the hypercubic lattice. 
The tight-binding band with a nearest-neighbor hopping has 
the nesting property at half-filling with the vector $\mib{Q}=(\pi, \cdots, \pi)$. 
The nesting may give rise to instabilities of the Fermi surface resulting in, for example, magnetic order or CDW. 
In this section, we give a brief summary of properties of the infinite-dimensional hypercubic lattice, following refs.~\citen{Georges} and \citen{Muller-Hartmann}.

We consider a hypercubic lattice in dimension $d$ with the nearest-neighbor hopping $t$. 
We scale the hopping as $t=D/2\sqrt{2d}$ to be applicable to $d=\infty$. 
The energy dispersion is then given by
\begin{align}
	\epsilon_{\mib{k}} = -\frac{D}{\sqrt{2d}}\sum_{i=1}^{d} \cos k_i,
\end{align}
where we take the lattice constant unity. 
If 
the vector $\mib{k}$ satisfies $\epsilon_{\mib{k}}=0$, which is a condition of the half-filled Fermi surface, then the vector $\mib{k}+\mib{Q}$ also gives $\epsilon_{\mib{k}+\mib{Q}}=0$. 
Consequently, in any 
dimension, 
the hypercubic lattice exhibits the complete nesting property at half filling.

The density of states $\rho(\epsilon)= N_0^{-1} \sum_{\mib{k}} \delta(\epsilon-\epsilon_{\mib{k}})$ 
with $N_0$ being the number of lattice sites
can be evaluated by means of the Fourier transform.
In the infinite-dimensional limit, $d=\infty$, $\rho(\epsilon)$ becomes the Gaussian
\begin{align}
	\rho(\epsilon) = \frac{1}{D} \sqrt{\frac{2}{\pi}} {\rm e}^{-2\epsilon^2/D^2}.
\label{eq:dos_Gaussian}
\end{align}
Therefore, the energy band has no cutoff in infinite dimensions. 
We choose the unit $D=1$ in this paper.

\begin{figure}[tb]
	\begin{center}
	\includegraphics[width=\linewidth]{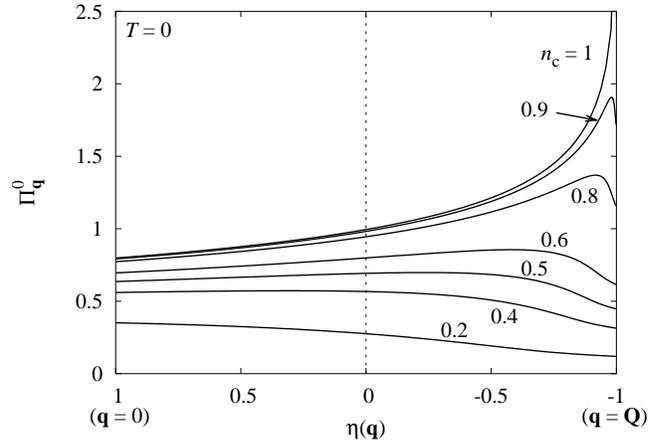}
	\end{center}
	\caption{
The polarization function 
$\Pi_{\mib{q}}^0$ of free conduction electrons at $T=0$ in the $d=\infty$ hypercubic lattice.}
	\label{fig:suscep_free}
\end{figure}

To see characteristics of 
responses in the hypercubic lattice, we evaluate the polarization function of free conduction electrons, $\Pi^0_{\mib{q}}$, defined by
\begin{align}
	\Pi^0_{\mib{q}} = \frac{1}{N_0} \sum_{\mib{k}}
	\frac{f_{\mib{k}+\mib{q}} - f_{\mib{k}}}{\epsilon_{\mib{k}} - \epsilon_{\mib{k}+\mib{q}}},
\end{align}
where $f_{\mib{k}}=1/(1+{\rm e}^{(\epsilon_{\mib{k}}-\mu)/T})$.
The staggered component at 
half filling  diverges logarithmically
with decreasing temperature due to the complete nesting, $\epsilon_{\mib{k}}=\epsilon_{\mib{k}+\mib{Q}}$. 
In the limit of $d=\infty$, the $\mib{q}$-dependence appears only through the quantity $\eta(\mib{q})$ defined by\cite{Muller-Hartmann}
\begin{align}
	\eta(\mib{q})= \frac{1}{d} \sum_{i=1}^d \cos q_i.
\end{align}
Figure~\ref{fig:suscep_free} shows filling dependences of $\Pi_{\mib{q}}^0$ at $T=0$.
At $n_{\rm c}=0.8$, a distinct maximum is found close to $\mib{q}=\mib{Q}$ ($\eta=-1$). 
At 
a density less than $n_{\rm c}=0.6$, on the other hand, the uniform component ($\eta=+1$) is favored rather than the staggered one ($\eta=-1$). 
%

Even taking the interaction into account,
the $\mib{q}$-dependence of the two-particle correlation is represented by $\eta(\mib{q})$ as long as the self-energy is local.
Especially for $\eta=+1$ and $-1$, the summation over $\mib{k}$ can be performed analytically.
In this paper, we compare the uniform and staggered components without considering possibility of incommensurate ordering.

\section{Properties at Half Filling}
\begin{figure}[tb]
	\begin{center}
	\includegraphics[width=\linewidth]{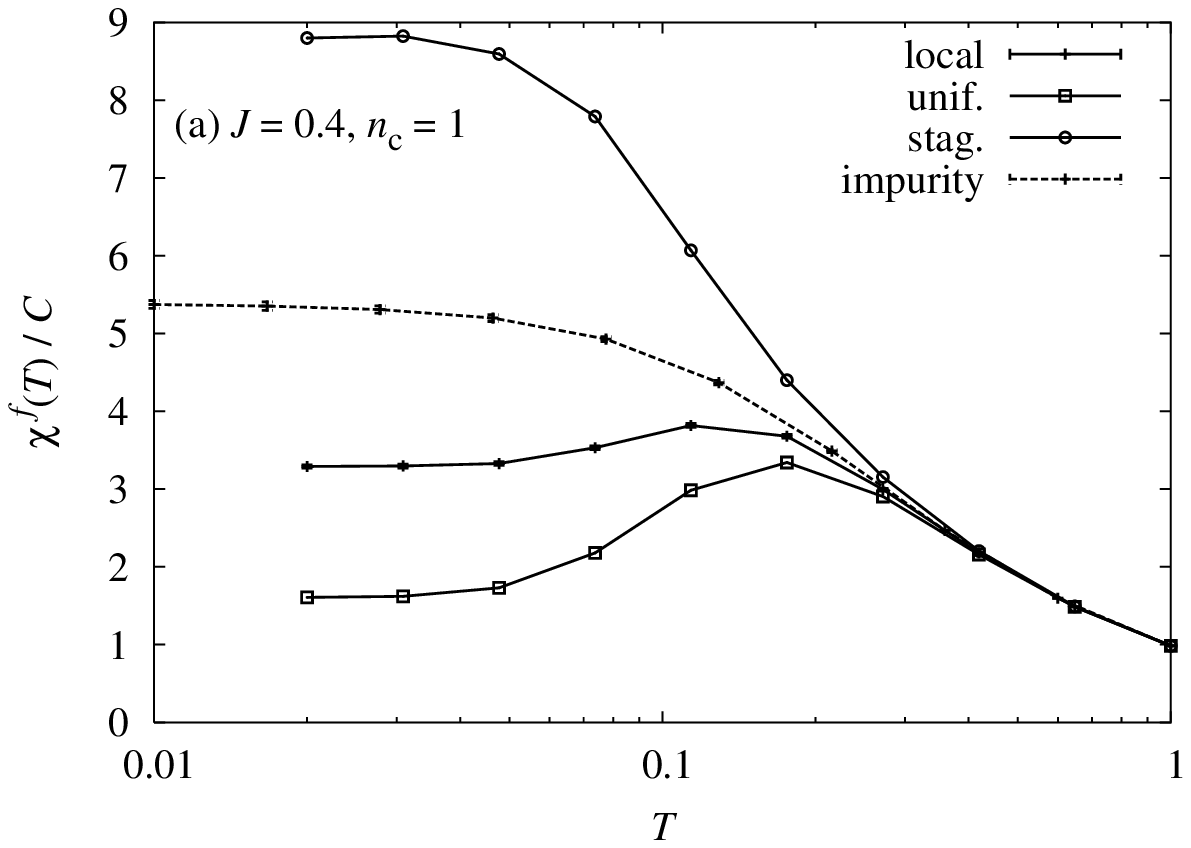}
	\includegraphics[width=\linewidth]{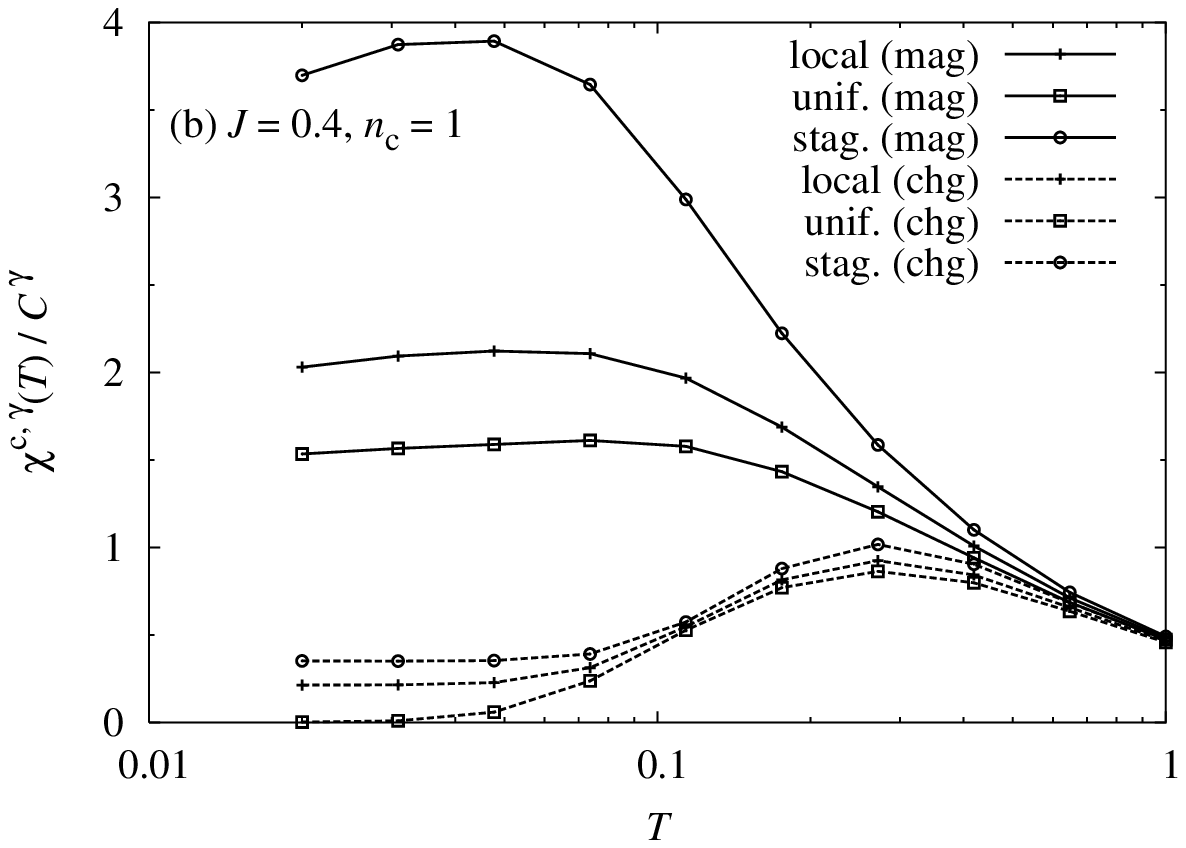}
	\end{center}
	\caption{
Temperature dependence of 
(a) the local $(\chi^f_{\rm loc})$, 
uniform $(\chi^f_{\mib{q}=0})$, 
and staggered $(\chi^f_{\mib{Q}})$ components of static susceptibilities of  local spins, 
and 
(b) the corresponding susceptibilities 
of conduction electrons.
The parameters are chosen as $J=0.4$ and $n_{\rm c}=1$. }
	\label{fig:KL-T_suscep1}
\end{figure}

We now present numerical results computed in the DMFT and the CT-QMC.
Let us begin with the half-filled band.
Figure~\ref{fig:KL-T_suscep1}(a) shows temperature dependences of the static susceptibilities of localized spins, $\chi^f_{\rm loc}$, $\chi^f_{\mib{q}=0}$ and $\chi^f_{\mib{Q}}$, for $J=0.4$.
For comparison, we also plot the susceptibility in the single-impurity system, $\chi_{\rm loc}^{\rm (imp)}$. 
At high temperatures, the susceptibilities simply exhibit local fluctuations following the Curie law, $\chi^f_{\mib{q}} \simeq C/T$ with $C=1/4$.
The local component $\chi^f_{\rm loc}$ takes a maximum at about $T=0.15$. 
This temperature dependence is contrastive to $\chi_{\rm loc}^{\rm (imp)}$, which monotonously increases with decreasing temperature. 
The reduction in the lattice system is due to a formation of an energy gap as shown later.
Compared to the local susceptibility, the staggered component is largely enhanced while the uniform one is suppressed. 
The large antiferromagnetic fluctuation is ascribed to the nesting of the Fermi surface. 

We show, in Fig.~\ref{fig:KL-T_suscep1}(b), the magnetic and charge susceptibilities of conduction electrons, $\chi^{{\rm c}, \gamma}$ $(\gamma=\text{mag}, \text{chg})$.
At high temperatures, the susceptibilities are approximately given by $\chi^{{\rm c}, \gamma} \simeq 2C^{\gamma} \Pi^0_{\mib{q}}$, where $C^{\gamma}$ corresponds to $C$ for the magnetic channel and unity for the charge channel.
The temperature dependence of magnetic susceptibilities is similar
to that of the localized spins for each case of $\chi^{{\rm c}, \rm{mag}}$, 
while the charge susceptibilities decrease at low temperatures.
In particular, the uniform charge fluctuation $\chi_{\mib{q}=0}^{\rm c, chg}$ eventually vanishes.  
This temperature dependence indicates formation of an energy gap, namely the Kondo insulator. 

\begin{figure}[tbp]
	\begin{center}
	\includegraphics[width=\linewidth]{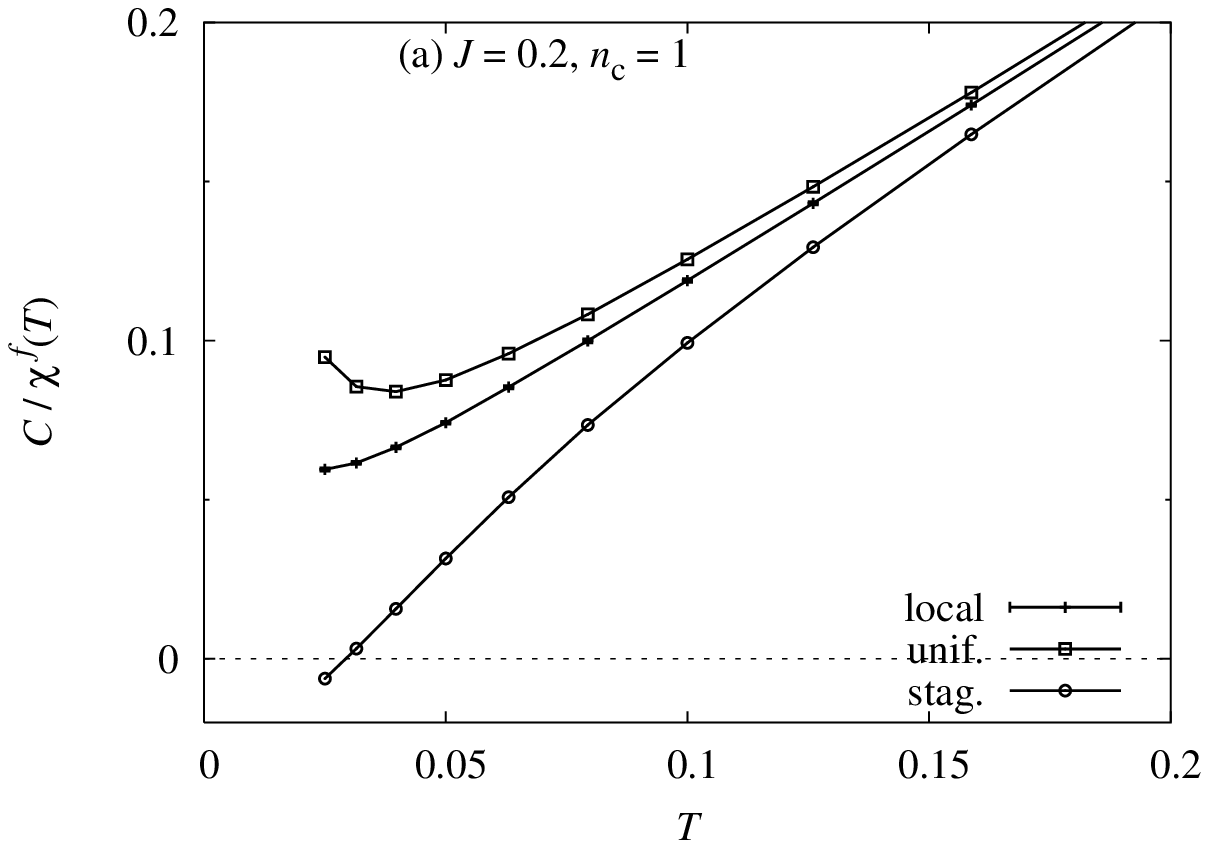}
	\includegraphics[width=\linewidth]{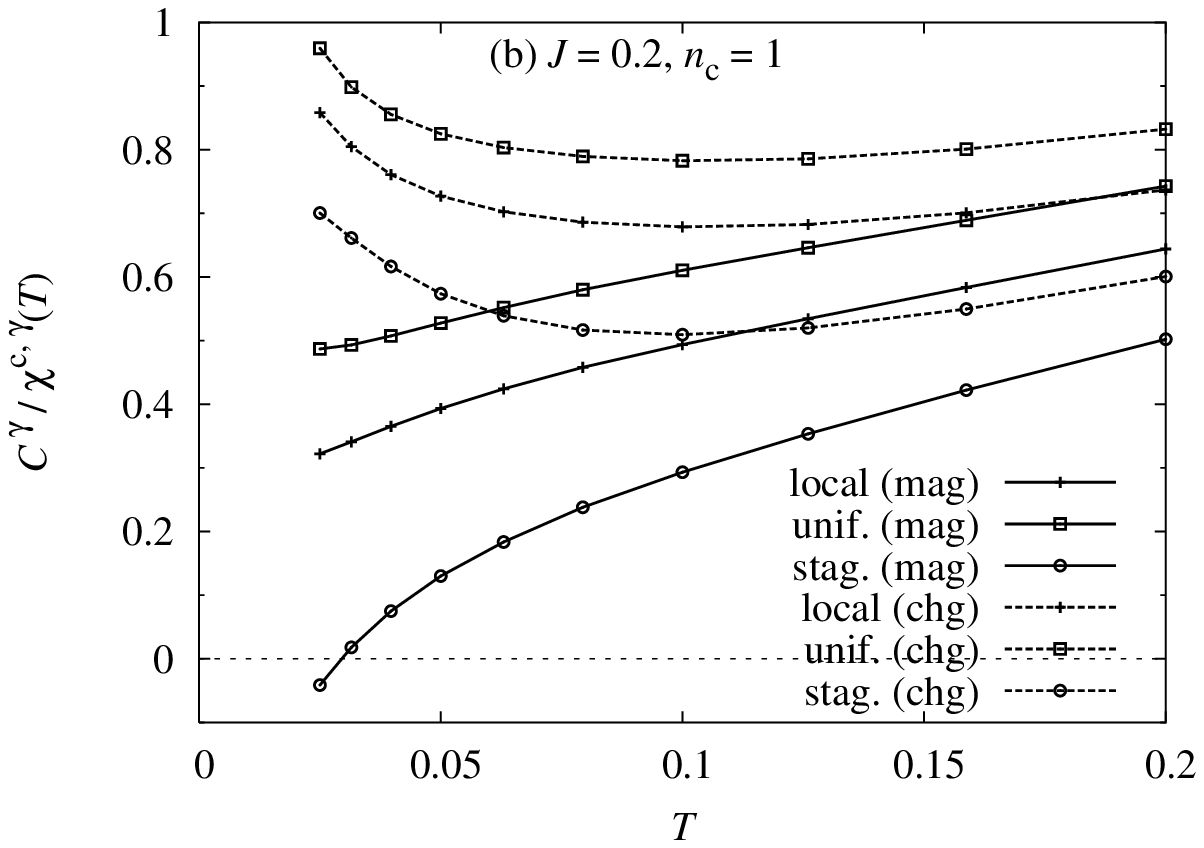}
	\end{center}
	\caption{
Temperature dependence of inverse static susceptibilities for $J=0.2$. }
	\label{fig:KL-T_suscep2}
\end{figure}

For smaller coupling, 
the characteristic energy becomes smaller with enhanced antiferromagnetic fluctuation.
The Kondo ground state then
becomes unstable against the RKKY interaction. 
Figure~\ref{fig:KL-T_suscep2}(a) shows temperature dependence of the inverse susceptibilities of the local spins for $J=0.2$ at $T \leq 0.2$.
We can recognize a divergence of the staggered magnetic susceptibility at $T_{\rm AF} \simeq 0.03$, indicating the antiferromagnetic transition. 
Since these calculations 
assume the paramagnetic effective medium at all temperatures, the negative susceptibility has been obtained below the transition temperature. 
In the present 
approximation based on the DMFT, the singularity in the susceptibility has no influence on the self-consistency of the effective medium. 

Near the transition temperature, the antiferromagnetic susceptibility 
follows the Curie-Weiss law, $\chi^f_{\mib{Q}}(T) / C \propto (T-T_{\rm AF})^{-1}$. 
This 
behavior originates in the mean-field treatment of the intersite interaction in the DMFT, 
which is justified in infinite dimensions.
The 
Curie-Weiss law has been reported also in the infinite-dimensional Hubbard model\cite{Jarrell-Hubbard}. 
The proportionality factor in the Curie-Weiss law is about 0.7 for $J=0.2$, which demonstrates a reduction of the local moment due to the Kondo effect. 
Figure~\ref{fig:KL-T_suscep2}(b) shows the corresponding conduction-electron susceptibilities. 
The staggered component of the magnetic channel diverges at $T_{\rm AF}$ 
affected by the corresponding fluctuation of the local spins. 
The charge fluctuations remain finite 
in contrast with the Kondo insulating state.

\begin{figure}[tbp]
	\begin{center}
	\includegraphics[width=\linewidth]{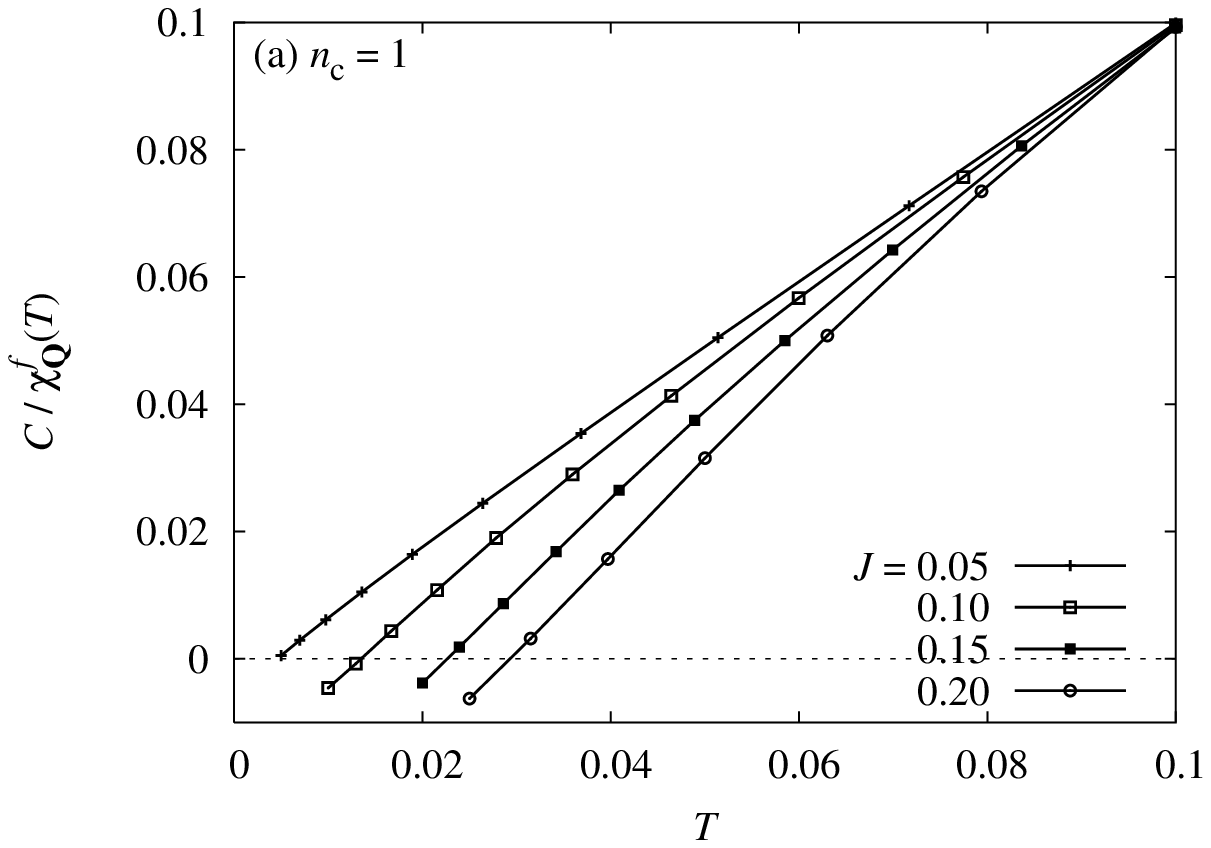}
	\includegraphics[width=\linewidth]{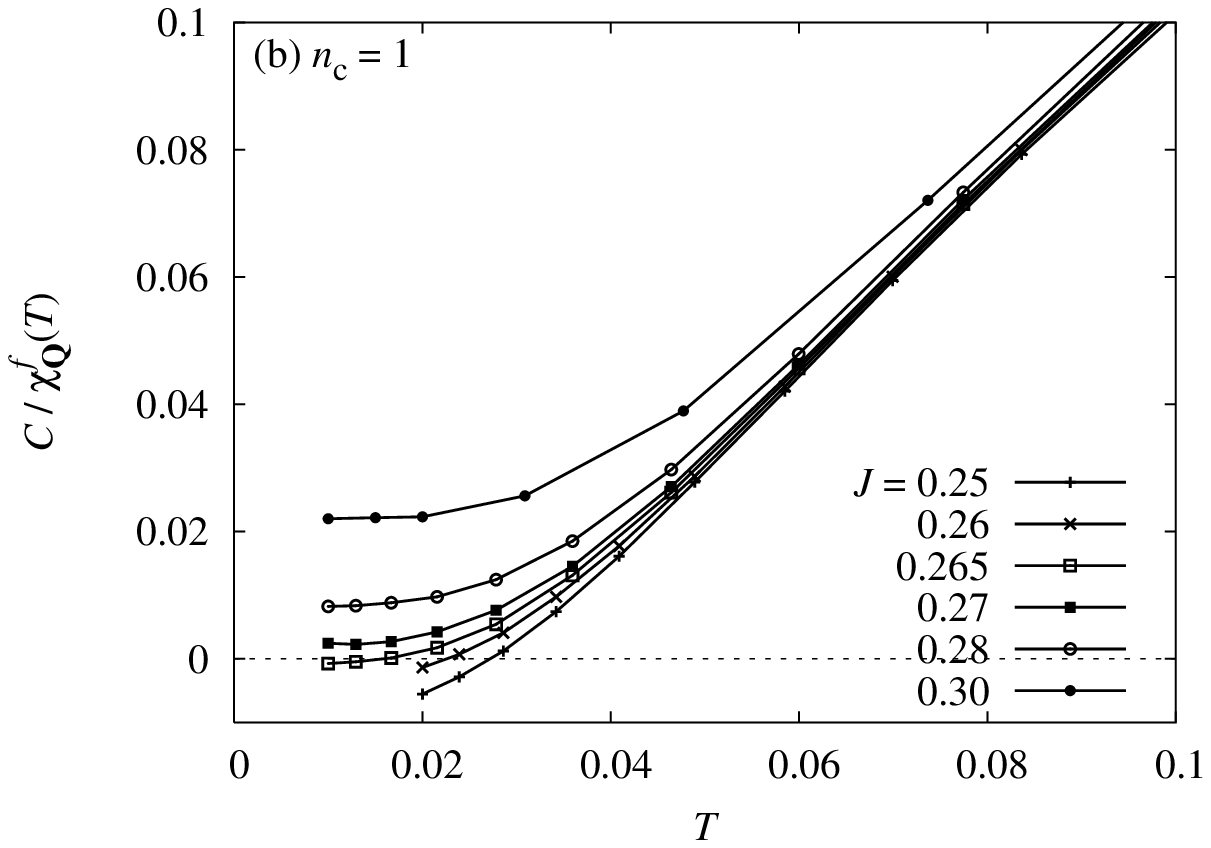}
	\end{center}
	\caption{
Temperature dependences of the inverse staggered susceptibility 
$1/\chi^f_{\mib{Q}}$ of localized spins for several values of $J$. }
	\label{fig:KL-T_suscep_stag_inv}
\end{figure}

To see the 
dependence of the transition temperature on $J$, we turn our attention to the staggered component $\chi^f_{\mib{Q}}$.
For 
small-$J$ case
such that $J \leq 0.2$ in Fig.~\ref{fig:KL-T_suscep_stag_inv}(a), the transition temperature $T_{\rm AF}$ monotonously increases with $J$.
The increased slope of $1/\chi^f_{\mib{Q}}$ indicates reduction of the moment. The reduction factor for the Curie constant changes from about 0.9 at $J=0.05$ to about 0.7 at $J=0.20$. 
As $J$ becomes larger 
than 0.2 in Fig.~\ref{fig:KL-T_suscep_stag_inv}(b), the susceptibility deviates from the Curie-Weiss law and tends to saturate.
Following the saturation of $\chi^f_{\mib{Q}}$, the transition temperature turns to decrease, and at $J=0.27$ we cannot see the divergence at finite temperatures.

\begin{figure}[tbp]
	\begin{center}
	\includegraphics[width=\linewidth]{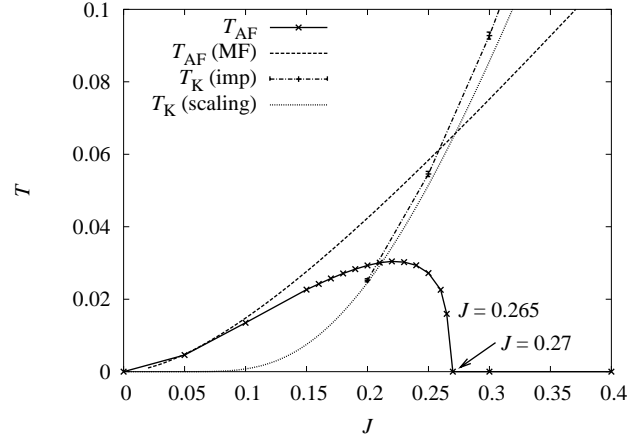}
	\end{center}
	\caption{Antiferromagnetic transition temperature $T_{\rm AF}$ as a function of $J$ at half filling. $T_{\rm AF}^{\rm (MF)}$ is the transition temperature evaluated in the mean-field theory. $T_{\rm K}^{\rm (imp)}$ and $T_{\rm K}^{\rm (scaling)}$ are the Kondo temperature determined from the low-temperature static susceptibility in the single-impurity system and from the formula in the scaling theory, eq.~(\ref{eq:TK_scaling2}), respectively. }
	\label{fig:KL-n050-Tc}
\end{figure}

We summarize the foregoing results on the antiferromagnetic instabilities at half filling.
Figure~\ref{fig:KL-n050-Tc} shows the antiferromagnetic transition temperature $T_{\rm AF}$ as a function of $J$. 
For comparison, we show the transition temperature $T_{\rm AF}^{\rm (MF)}$ evaluated in the mean-field theory, eq.~(\ref{eq:Tc_mf}).
At weak coupling of $J=0.05$, the mean-field theory turns out to reproduce the DMFT results quantitatively. 
This follows from the fact that the DMFT deals with the intersite interactions as the mean field. 
As $J$ increases, the transition temperature in the DMFT falls below the mean-field results owing to the development of the local correlation, which screens the local spins from the surrounding spins. 
The maximum of $T_{\rm AF}$ is found to be about 0.03 at $J \simeq 0.22$.
After taking the maximum, $T_{\rm AF}$ rapidly 
decreases around $J \simeq 0.25$ and eventually vanishes at $J_{\rm c} \simeq 0.27$.

Following Doniach\cite{Doniach}, we estimate the critical value of the coupling constant by comparing the Kondo temperature $T_{\rm K}$ and the RKKY interaction. 
We invoke the single-impurity system to evaluate $T_{\rm K}$. 
From the static susceptibility $\chi^{\rm (imp)}$ at zero temperature, 
$T_{\rm K}$ is defined by $T_{\rm K}^{\rm (imp)} = C/\chi^{\rm (imp)}$.
We also estimate the Kondo temperature with use of the following analytic expression derived by means of the scaling theory:\cite{Hewson}
\begin{align}
	T_{\rm K}^{\rm (scaling)} = D\sqrt{2J \rho(\mu)} \exp \left[
	-\frac{1}{2J \rho(\mu)} \right], 
\label{eq:TK_scaling2}
\end{align}
where $\mu$ denotes the chemical potential for $J=0$.
While this expression is originally derived for the constant density of states, 
eq.~(\ref{eq:TK_scaling2}) holds well also for the Gaussian density of states
as long as $T_{\rm K}$ is much smaller than the band width. 
The Kondo temperature is plotted in Fig.~\ref{fig:KL-n050-Tc} together with the mean-field transition temperature $T_{\rm AF}^{\rm (MF)}$ evaluated from eq.~(\ref{eq:Tc_mf}).
It turns out that $T_{\rm AF}^{\rm (MF)}$ crosses $T_{\rm K}^{\rm (imp)}$ and $T_{\rm K}^{\rm (scaling)}$ between $J=0.25$ and $J=0.30$. 
Hence, 
comparison of 
the energy scales reproduces fairly well the critical coupling constant. These numerical results support 
Doniach's picture for competition between the Kondo singlet and the magnetic ordering due to the RKKY interaction. 

As shown in Fig.~\ref{fig:KL-T_suscep_stag_inv}(b), the staggered susceptibility around the critical coupling $J_{\rm c}$ saturates at low temperatures.
Hence, the expansion of $\chi^f_{\mib{Q}}(T)^{-1}$ with respect to $T$ does not include a linear term near $J_{\rm c}$. 
We approximate this behavior as
\begin{align}
	\chi_{\mib{Q}}(T, J)^{-1} = a T^2 + b (J-J_{\rm c}).
\end{align}
Equating $\chi_{\mib{Q}}(T_{\rm AF}, J)^{-1}$ to 0, we obtain the transition temperature near $J_{\rm c}$ as
\begin{align}
	T_{\rm AF}(J) \propto (J_{\rm c}-J)^{1/2}.
\end{align}
Consequently, the 
slope of $T_{\rm AF}(J)$ diverges at $J_{\rm c}$. 
This feature can be 
seen in Fig.~\ref{fig:KL-n050-Tc}.


\section{Doping on Insulating States}

We have so far examined the half filling case $n_{\rm c}=1$, and have 
observed two insulators: the antiferromagnetic insulator 
and the Kondo insulator. 
Let us next examine the filling-dependence of these two phases. 
We expect reduction of the antiferromagnetic fluctuation, and crossover to the metallic state from the Kondo insulator.
Because of the particle-hole symmetry of the Kondo coupling, $n_{\rm c}$ and $2-n_{\rm c}$ are 
equivalent.
Henceforth, we investigate hole-doping 
$n_{\rm c}<1$ only.

\begin{figure}[tbp]
	\begin{center}
	\includegraphics[width=\linewidth]{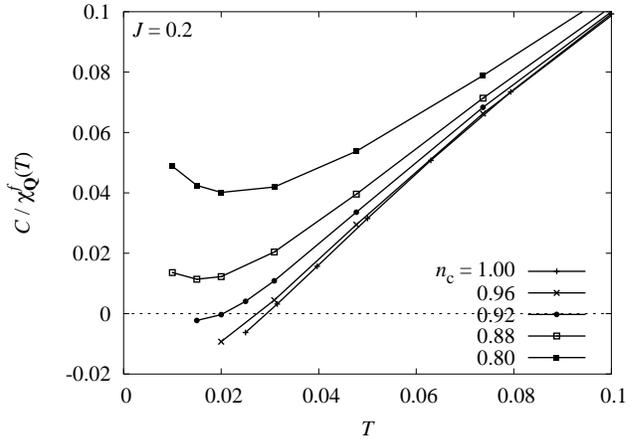}
	\end{center}
	\caption{
Temperature dependence of the inverse staggered susceptibility $1/\chi^f_{\mib{Q}}$ of localized spins near half filling with $J=0.2$.}
	\label{fig:KL-T_suscep_near_n050}
\end{figure}
We first examine the antiferromagnetic phase. 
Figure~\ref{fig:KL-T_suscep_near_n050} shows temperature dependences of the inverse of the staggered susceptibility at $J=0.2$ for several values of $n_{\rm c}$ near half filling. 
Doping on the half-filled band suppresses the 
antiferromagnetic fluctuation. 
This is due to the change of 
the nesting condition in the Fermi surface.
At $n_{\rm c}=0.88$, the antiferromagnetic transition disappears, 
and the paramagnetic ground state is stabilized. 
\begin{figure}[tbp]
	\begin{center}
	\includegraphics[width=\linewidth]{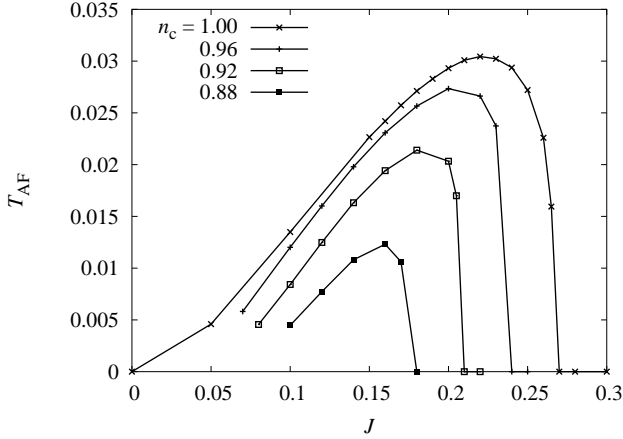}
	\end{center}
	\caption{Antiferromagnetic transition temperature $T_{\rm AF}$ as a function of $J$ for several values of $n_{\rm c}$. }
	\label{fig:KL-n-T_AF}
\end{figure}
In Fig.~\ref{fig:KL-n-T_AF} we show an antiferromagnetic 
transition temperature near half filling. 
The phase boundaries seem to be almost scaled
by $J$.
At $n_{\rm c}=0.80$, we have observed no transition down to $T=0.005$.

\begin{figure}[tb]
	\begin{center}
	\includegraphics[width=\linewidth]{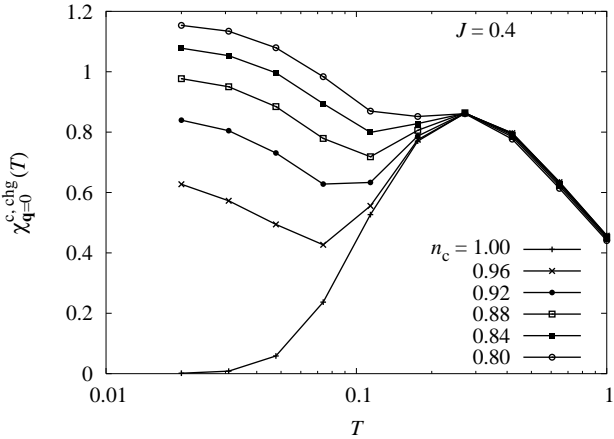}
	\end{center}
	\caption{Temperature dependence of the uniform charge susceptibility $\chi^{\rm c, chg}_{\mib{q}=0}$ of conduction electrons 
near half filling
for $J=0.4$.}
	\label{fig:KL-T_suscep_c_unif_near_n050}
\end{figure}
We turn our attention to the Kondo-insulator. 
Figure~\ref{fig:KL-T_suscep_c_unif_near_n050} shows filling dependences of the uniform charge susceptibility, $\chi^{\rm c, chg}_{\mib{q}=0}$, for $J=0.4$. 
The charge susceptibility 
is sensitive to 
the doping, which gives rise to particle-hole excitations and a 
finite charge susceptibility at absolute zero.
Similar 
non-monotonous behavior of $\chi^{\rm c, chg}_{\mib{q}=0}(T)$ has been observed in the one-dimensional Kondo lattice\cite{Shibata-Tsunetsugu}.
These two energy scales also appear in the single-particle excitation spectrum, which shows an energy gap of order $T_{\rm K}$ and the coherent state at low energies\cite{Otsuki-LFS}.


\section{Ferromagnetism at Low Carrier Densities}
At low density, the conduction electrons do not have the nesting property.
The polarization function $\Pi_{\mib{q}}^0$ in Fig.~\ref{fig:suscep_free} 
exhibits a flat structure at wide range of $\mib{q}$.
The maximum occurs at $\mib{q}=0$ rather than $\mib{Q}$.  
The RKKY interaction accordingly favors ferromagnetic ordering rather than antiferromagnetism at low carrier density.

\begin{figure}[tbp]
	\begin{center}
	\includegraphics[width=\linewidth]{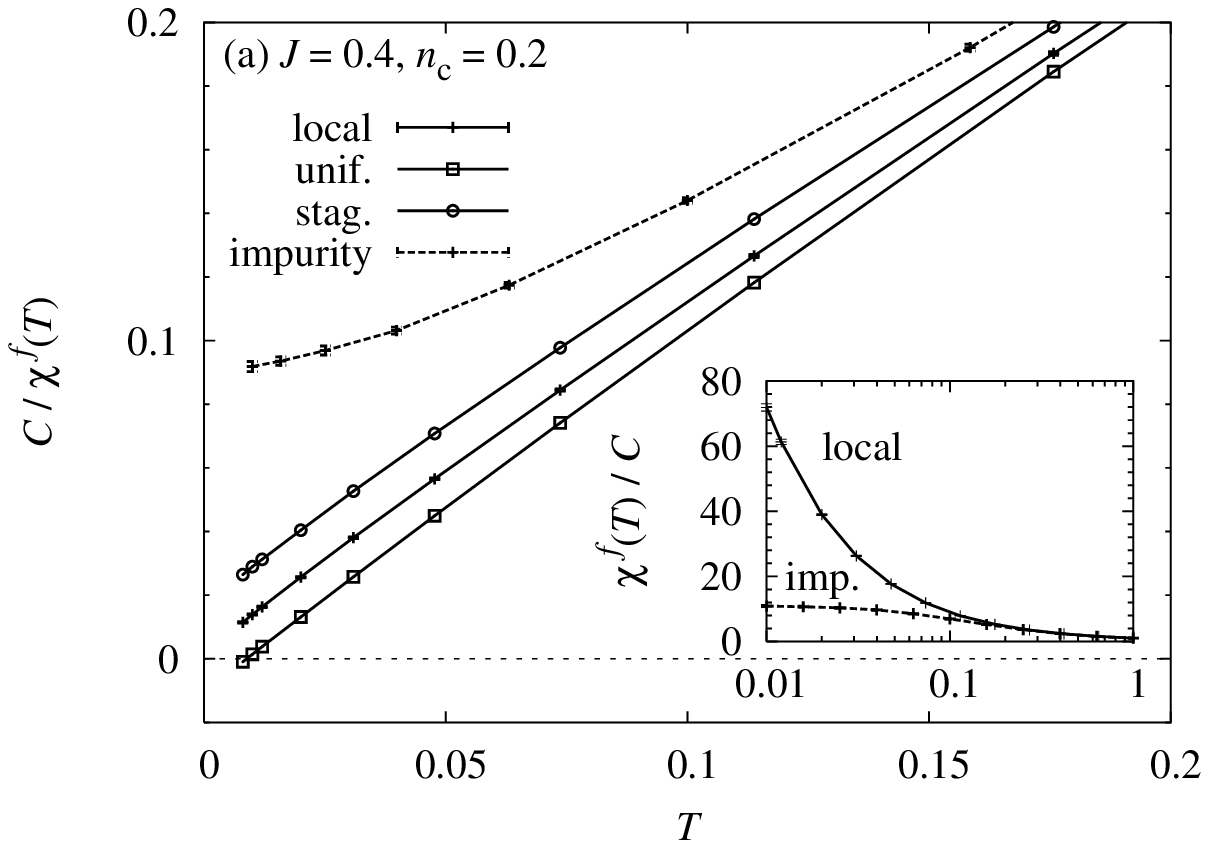}
	\includegraphics[width=\linewidth]{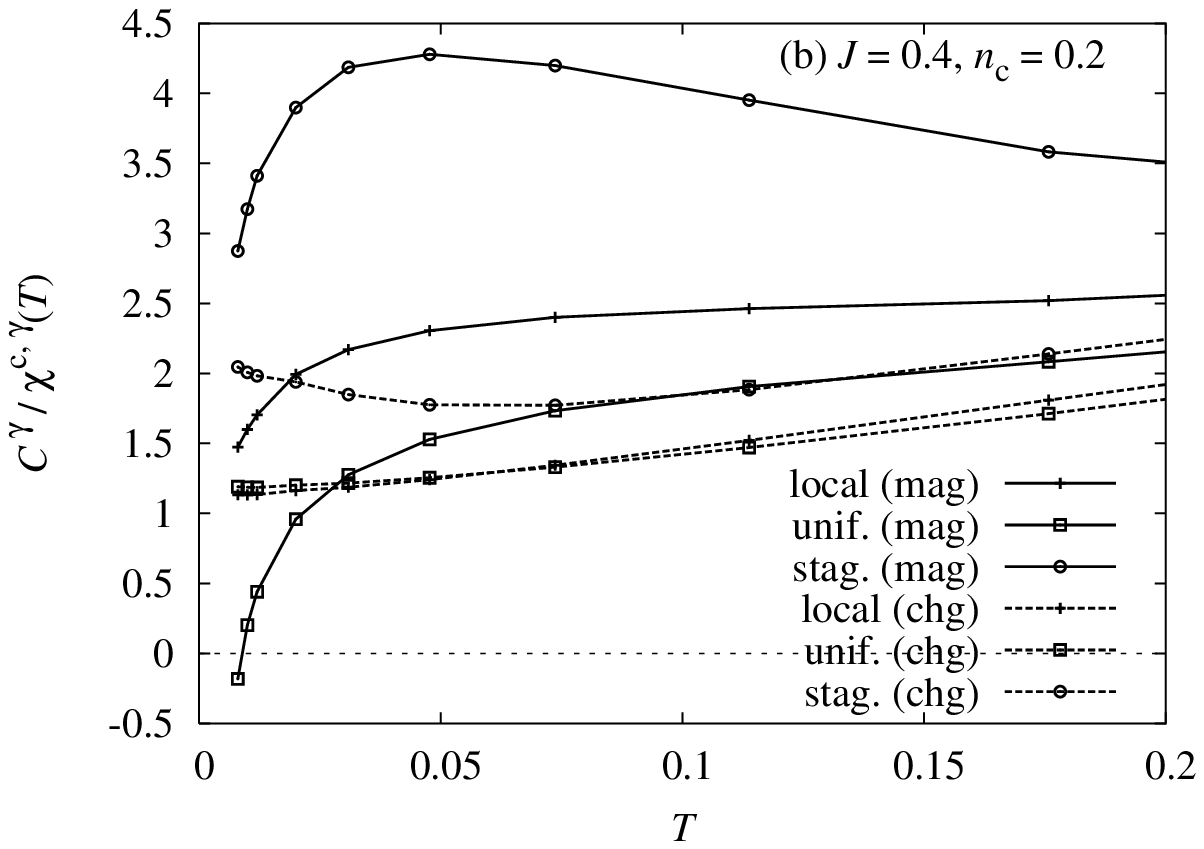}
	\end{center}
	\caption{Temperature dependence of inverses static susceptibilities
for $J=0.4$ and $n_{\rm c}=0.2$. }
	\label{fig:KL-n010-T_suscep_inv}
\end{figure}

Let us compare the local susceptibilities in the single-impurity and periodic systems. 
If the number of conduction electrons are less than the number of local spins,
it is impossible to screen all the spins even in the strong coupling limit $J=\infty$. 
This is in strong contrast with the impurity system.
Figure~\ref{fig:KL-n010-T_suscep_inv}(a) shows temperature dependence of inverse magnetic susceptibilities for $J=0.4$ and $n_{\rm c}=0.2$. 
The inset shows that the susceptibility in 
the periodic system continues to grow in the temperature range where
the impurity susceptibility is saturated.
It turns out 
from Fig.~\ref{fig:KL-n010-T_suscep_inv}(a) 
that the susceptibility of the Kondo lattice follows the Curie-Weiss law 
in
a wide temperature range, which
indicates an existence of the almost free moment.
In contrast, the impurity model has a large deviation from the Curie-Weiss law due to the Kondo screening.

As expected from the behavior of $\Pi_{\mib{q}}^0$ in Fig.~\ref{fig:suscep_free}, the uniform component is larger than the staggered one. 
The uniform susceptibility eventually diverges at a finite temperature, indicating the ferromagnetic transition. 
The corresponding susceptibility $\chi^{\rm c}$ of conduction electrons is shown in Fig.~\ref{fig:KL-n010-T_suscep_inv}(b).
The ferromagnetic component diverges following the behavior of local spins. 

\begin{figure}[tbp]
	\begin{center}
	\includegraphics[width=\linewidth]{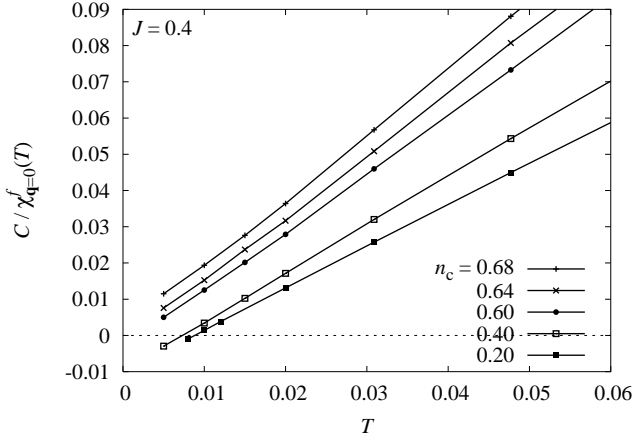}
	\end{center}
	\caption{Temperature dependence of inverse uniform susceptibility $C/\chi^f_{\mib{q}=0}$ for $J=0.4$ and several values of $n_{\rm c}$. }
	\label{fig:KL-J040-T_suscep_unif_inv}
\end{figure}
\begin{figure}[tb]
	\begin{center}
	\includegraphics[width=\linewidth]{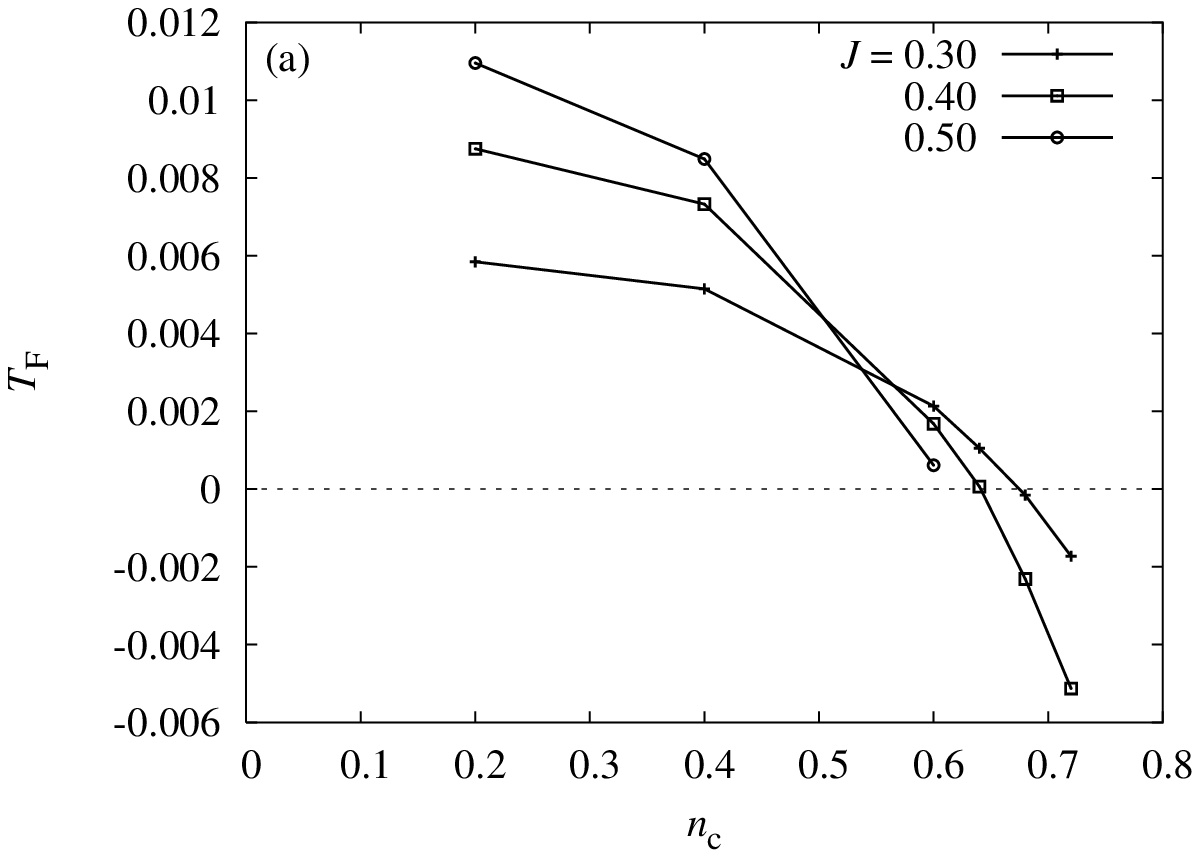}
	\includegraphics[width=\linewidth]{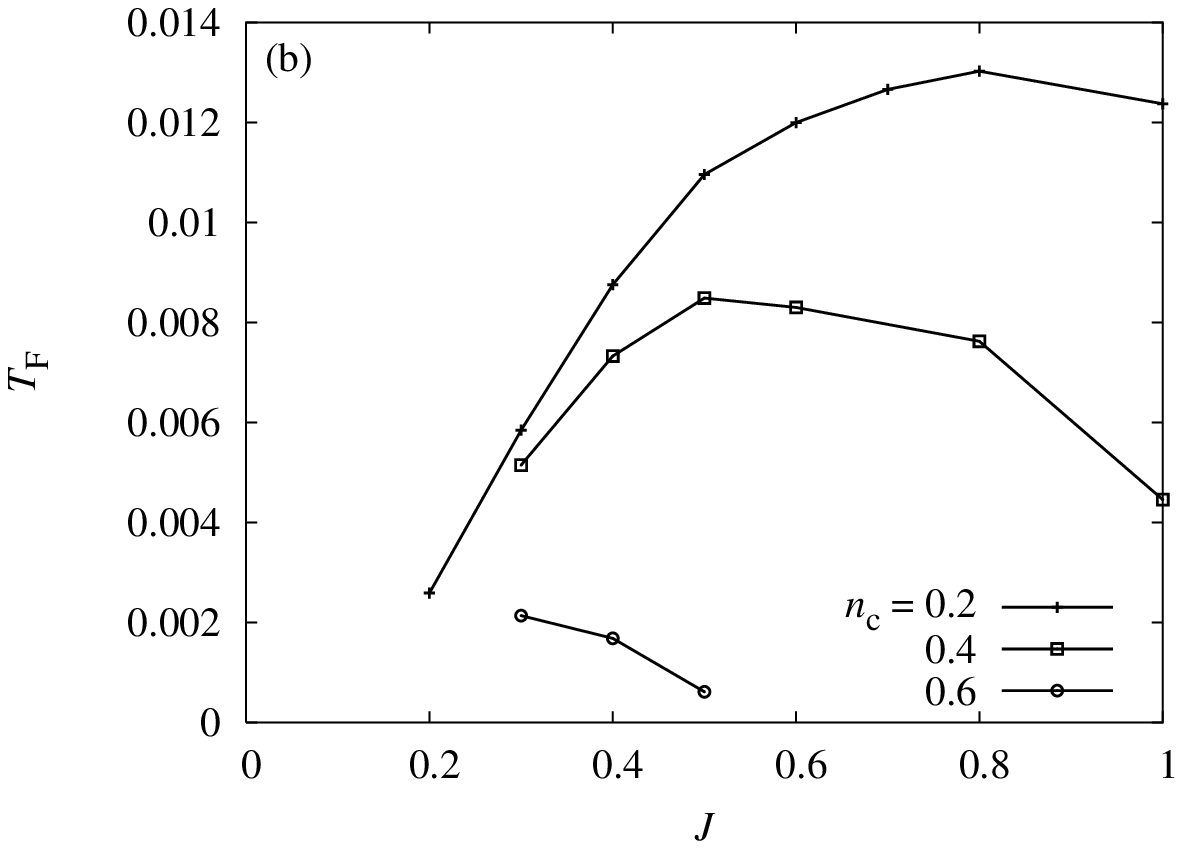}
	\end{center}
	\caption{Ferromagnetic transition temperature $T_{\rm F}$ (a) as a function of $n_{\rm c}$ for $J=0.3$, 0.4 and 0.5, and (b) as a function $J$ for $n_{\rm c}=0.2$, 0.4 and 0.6. }
	\label{fig:KL-T_F}
\end{figure}

Let us examine the filling-dependence of the ferromagnetic transition temperature at low carrier densities. 
Figure~\ref{fig:KL-J040-T_suscep_unif_inv} shows the inverse uniform susceptibility for various values of $n_{\rm c}$ up to 0.68 with $J=0.4$. 
We can clearly see suppression of the ferromagnetic correlation with increasing 
$n_{\rm c}$.
We estimate the ferromagnetic transition temperatures $T_{\rm F}$ by 
extrapolating the data at $T=0.005$ and 0.010 according to the Curie-Weiss law. 
At the low-carrier-density regime, such an extrapolation is expected to be applicable reasonably. 
Figure~\ref{fig:KL-T_F}(a) shows $T_{\rm F}$ as a function of $n_{\rm c}$ for $J=0.3$, 0.4 and 0.5.
As $n_{\rm c}$ increases, $T_{\rm F}$ decreases 
and turn to negative values. 
For $J=0.4$, the ferromagnetic transition disappears at about $n_{\rm c}=0.64$. 
%
%
%

Figure~\ref{fig:KL-T_F}(b) shows the transition temperature $T_{\rm F}$ as a function of $J$ at $n_{\rm c}=0.2$, 0.4 and 0.6.
After taking a maximum against $J$,
$T_{\rm F}$ decreases gradually, in contrast with the antiferromagnetic case shown in Figs.~\ref{fig:KL-n050-Tc} and \ref{fig:KL-n-T_AF}. 
At $n_{\rm c}=0.2$, 
the ferromagnetism persists beyond $J=1.3$.
If we compare $T_{\rm K}$ in eq.~(\ref{eq:TK_scaling2}) and $T_{\rm F}^{\rm (MF)}$ in eq.~(\ref{eq:Tc_mf}), we obtain the critical coupling constant as $J\simeq 0.6$ for $n_{\rm c}=0.2$. 
Hence, Doniach's picture does not hold 
in the low-carrier-density regime, where the carrier number is insufficient to screen all local spins.


\section{CDW at Quarter Filling}
In this section, we show results for quarter filling $n_{\rm c}=0.5$.
\begin{figure*}[tb]
	\begin{center}
	\includegraphics[width=0.49\linewidth]{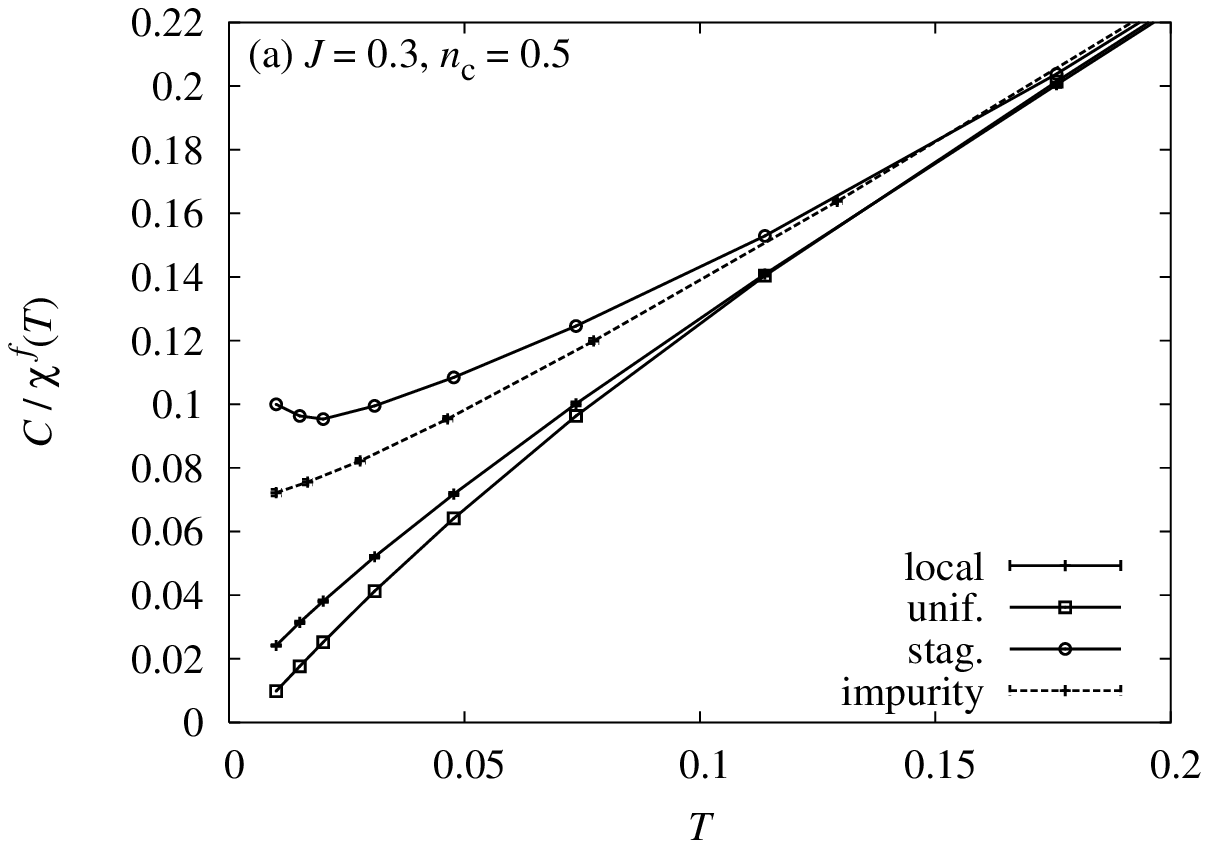}
	\includegraphics[width=0.49\linewidth]{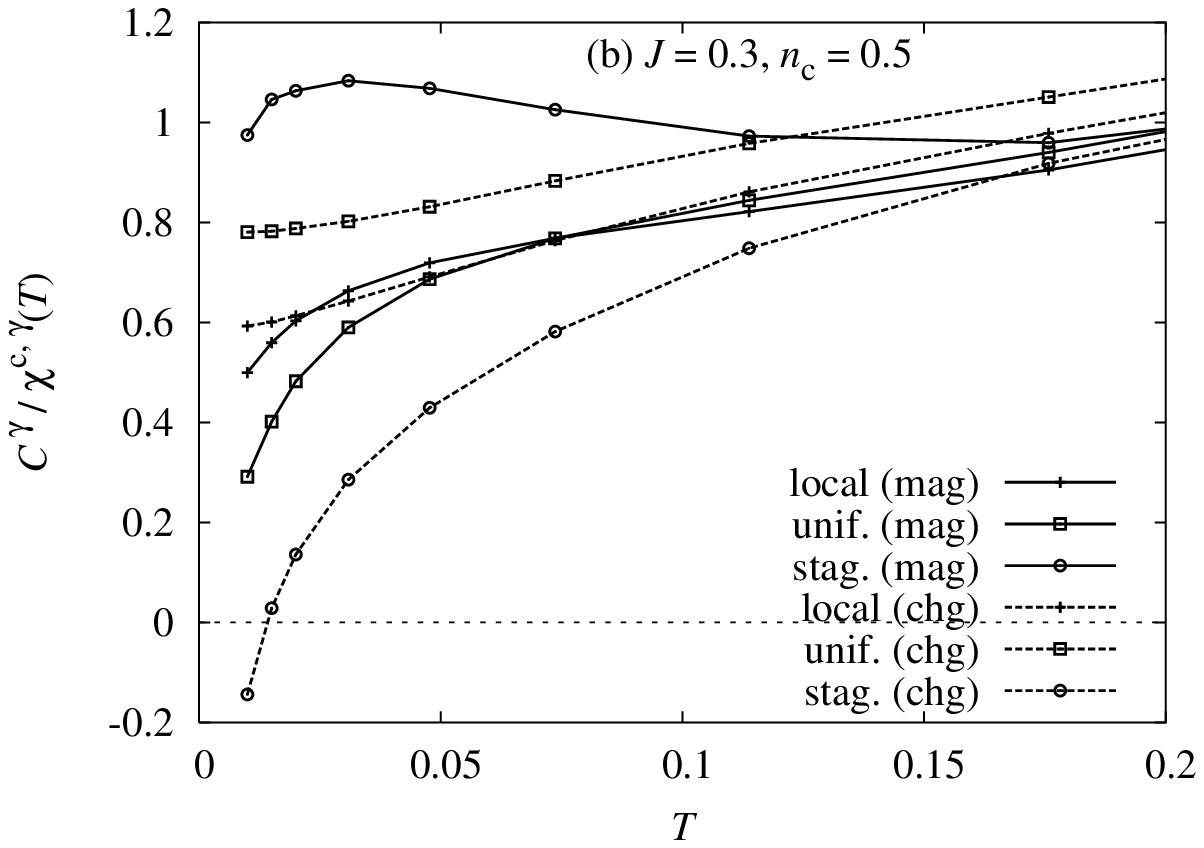}
	\includegraphics[width=0.49\linewidth]{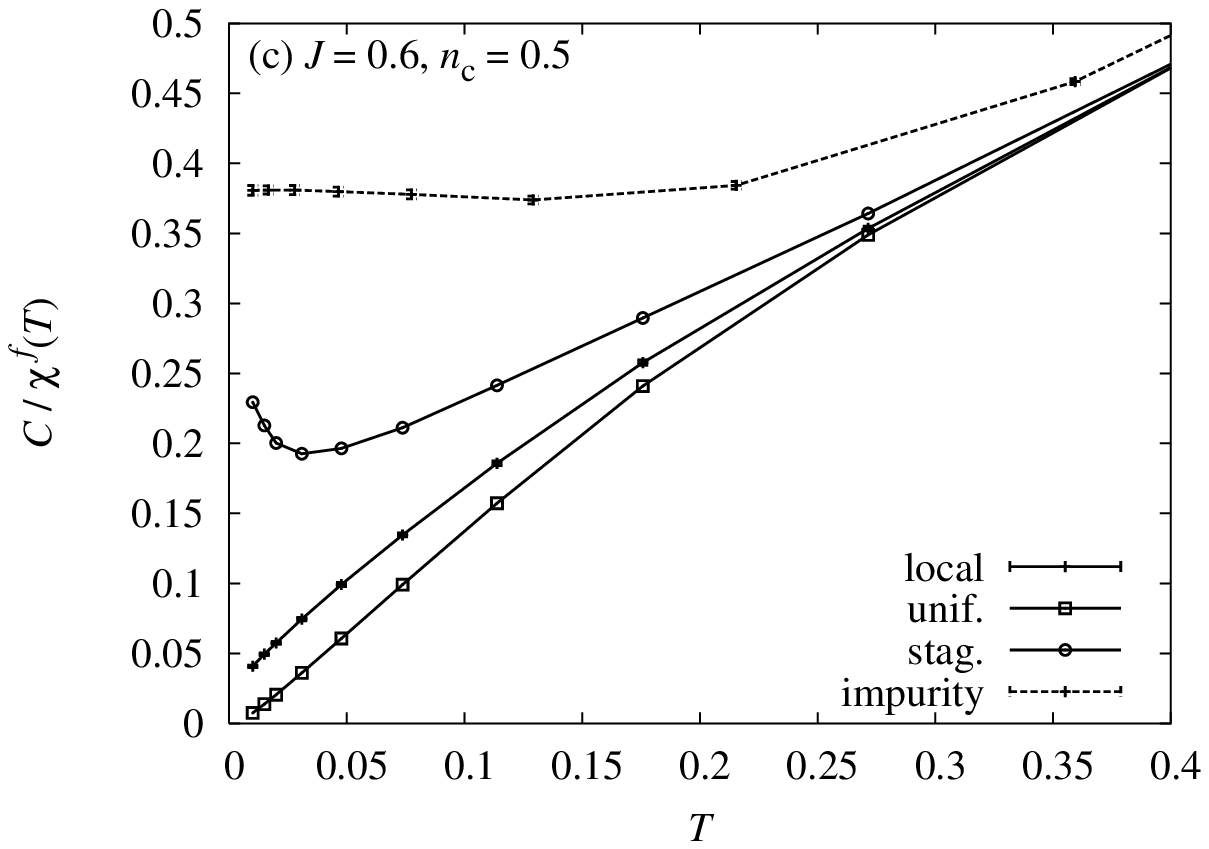}
	\includegraphics[width=0.49\linewidth]{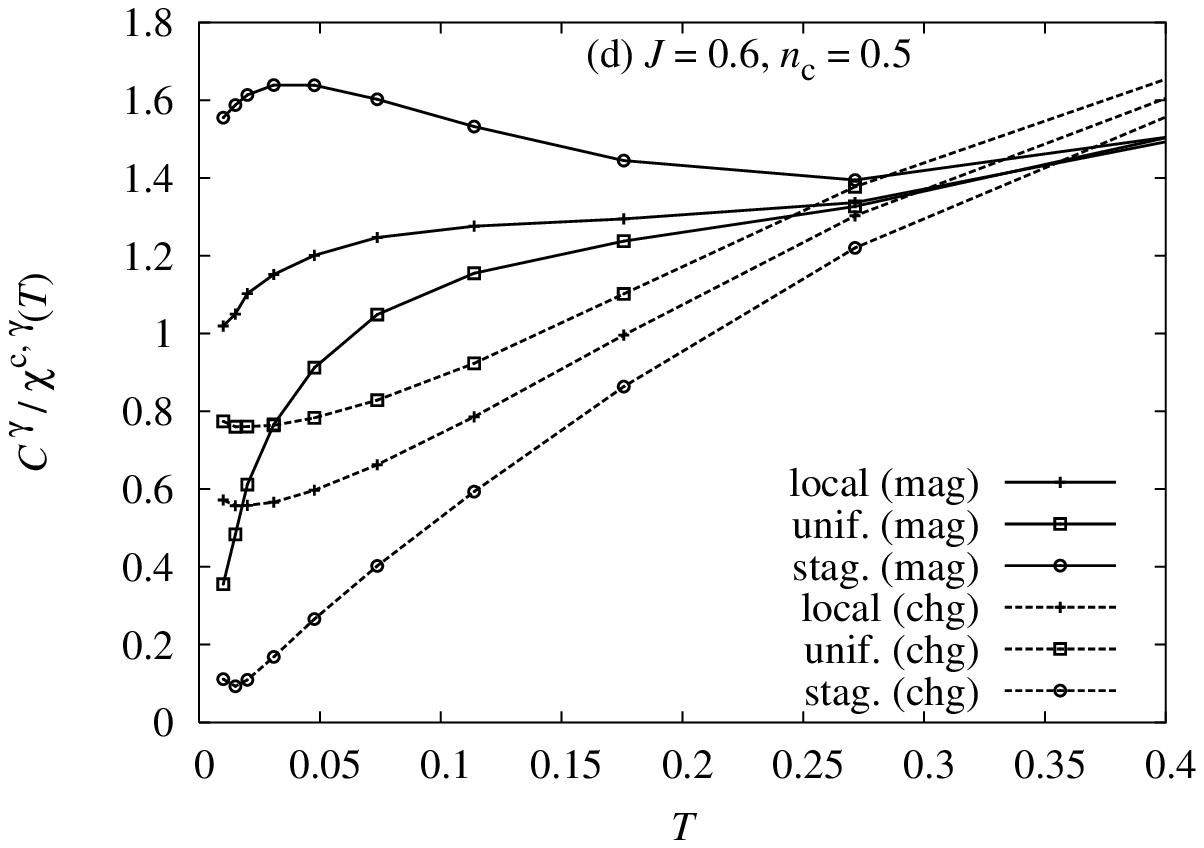}
	\end{center}
	\caption{Temperature dependence of the inverse static susceptibilities at quarter filling for $J=0.3$ in (a) and (b), and $J=0.6$ in (c) and (d). }
	\label{fig:KL-J030-n025-T_suscep_q}
\end{figure*}
%
%
Figure~\ref{fig:KL-J030-n025-T_suscep_q}(a) shows inverse magnetic susceptibilities of localized 
spins at $J=0.3$. 
The local spins have the ferromagnetic correlation.
An extrapolation reveals that the divergence of the ferromagnetic fluctuation takes place at the temperature of order $10^{-3}$. 

We show the corresponding conduction-electron susceptibilities in Fig.~\ref{fig:KL-J030-n025-T_suscep_q}(b). 
The staggered component of the charge susceptibility $\chi_{\mib{Q}}^{\rm c, chg}$ diverges at about $T \simeq 0.014$, 
indicating an ordering into the charge-density-wave (CDW) state. 
Concerning the magnetic channel, the uniform component is favored following the corresponding fluctuation of the local spins. 
For stronger coupling constant, $J=0.6$, 
The ferromagnetic correlation remains toward divergence (Fig.~\ref{fig:KL-J030-n025-T_suscep_q}(c)). 
On the other hand, the increase of the CDW fluctuation is suppressed at low temperatures as shown in Fig.~\ref{fig:KL-J030-n025-T_suscep_q}(d). 

\begin{figure}[tbp]
	\begin{center}
	\includegraphics[width=\linewidth]{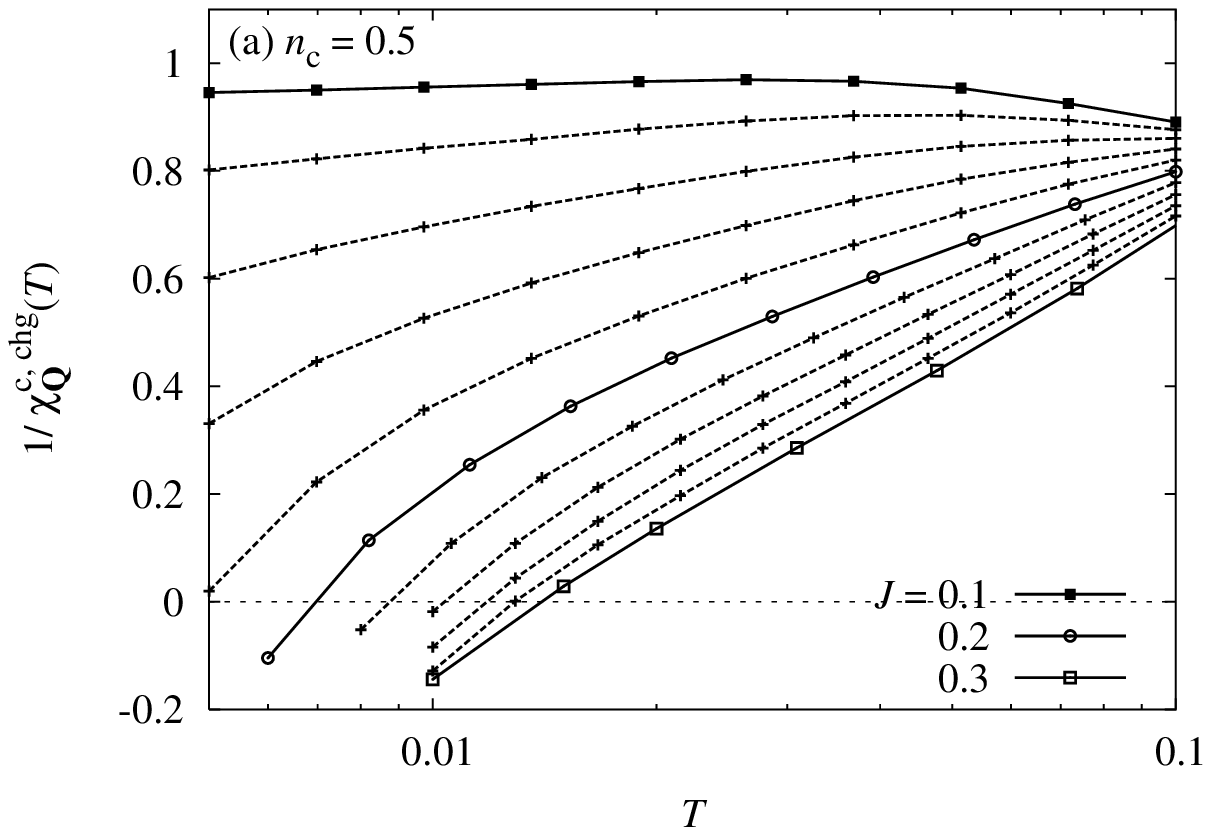}
	\includegraphics[width=\linewidth]{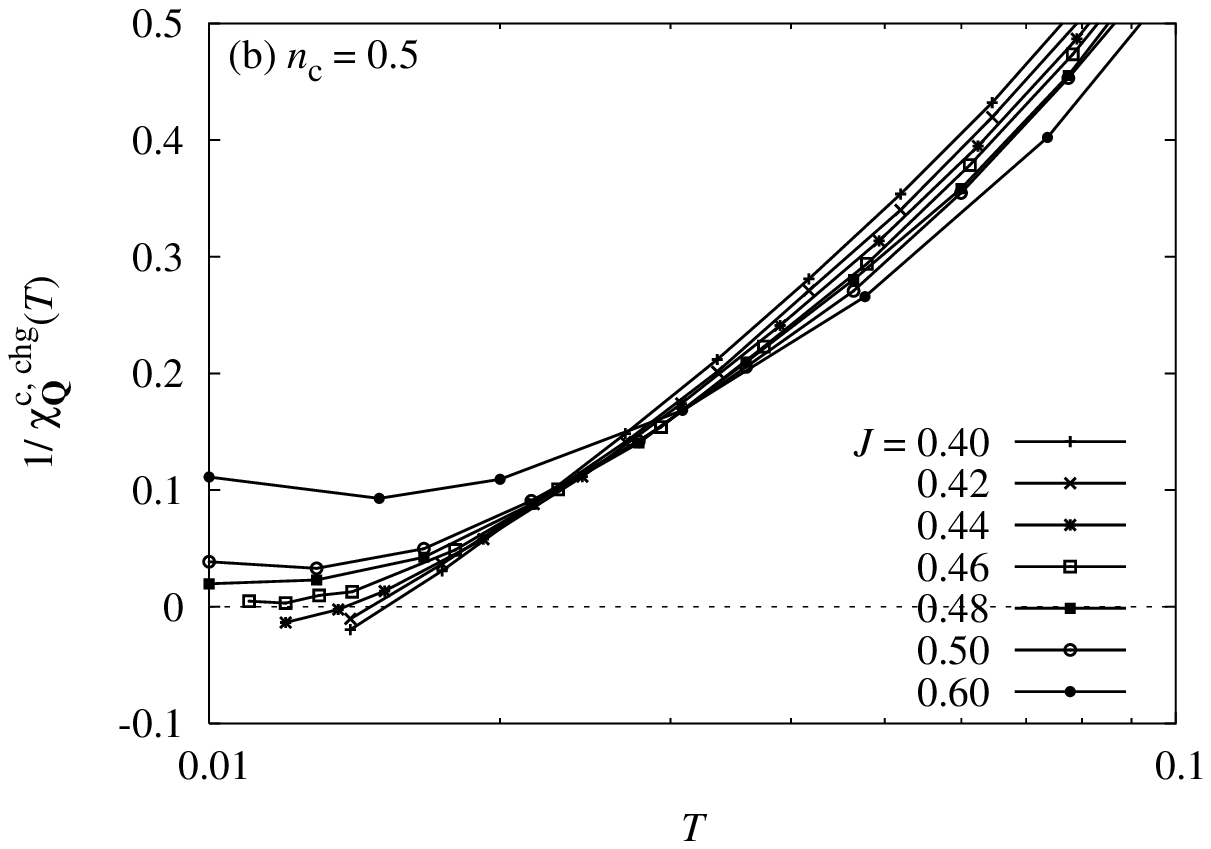}
	\end{center}
	\caption{Temperature dependence of inverse 
staggered charge susceptibilities $1/\chi^{\rm c, chg}_{\mib{Q}}$ for conduction electrons exhibiting the CDW instability. 
The parameter $J$ varies by 0.02 at every step to an adjacent curve.}
	\label{fig:KL-n025-T_suscep_c_stag}
\end{figure}

\begin{figure}[tbp]
	\begin{center}
	\includegraphics[width=\linewidth]{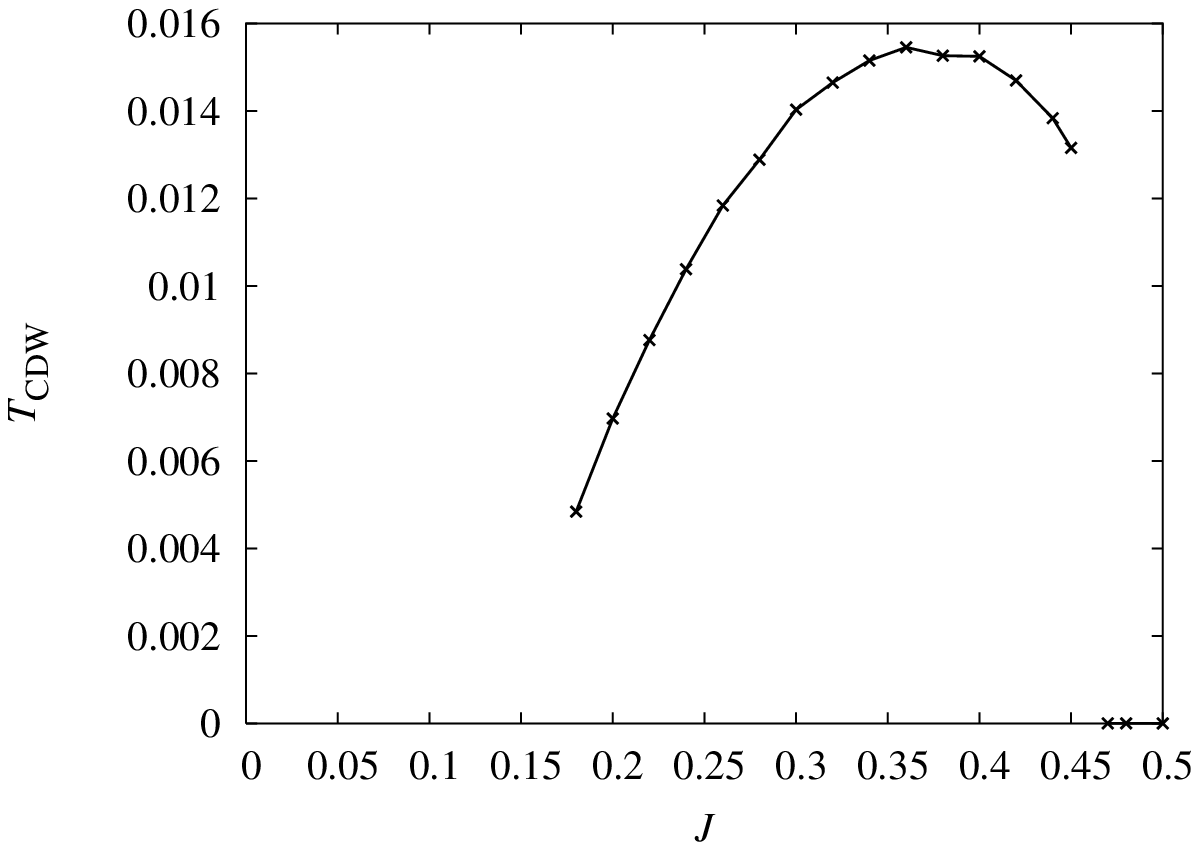}
	\end{center}
	\caption{The CDW ($\mib{q}=\mib{Q}$) transition temperature $T_{\rm CDW}$ as a function of $J$ for the quarter filling $n_{\rm c}=0.5$. }
	\label{fig:KL-n025-tc}
\end{figure}

We concentrate our attention on the CDW 
instability.
We first vary $J$ keeping $n_{\rm c}$.
Figure~\ref{fig:KL-n025-T_suscep_c_stag} shows $1/\chi_{\mib{Q}}^{\rm c, chg}$ as a function of temperature in logarithmic scale.
As $J$ decreases, the transition temperature decreases, and for $J<0.18$ no divergence is found down to $T=0.005$. 
However, 
we cannot exclude the CDW ordering at these parameters, since the energy scale seems to decrease exponentially against $J$ as shown in Fig.~\ref{fig:KL-n025-T_suscep_c_stag}(a). 
On the other hand, a boundary on the strong-coupling side exists as is clear from Fig.~\ref{fig:KL-n025-T_suscep_c_stag}(b). 
The susceptibility tends to saturate with increasing $J$. 
The CDW transition temperature $T_{\rm CDW}$ is plotted in Fig.~\ref{fig:KL-n025-tc}. 
The maximum of $T_{\rm CDW}$ is about 0.016 at 
$J \sim 0.36$. 
The CDW transition is completely suppressed at $J \geq 0.47$.

\begin{figure}[tbp]
	\begin{center}
	\includegraphics[width=\linewidth]{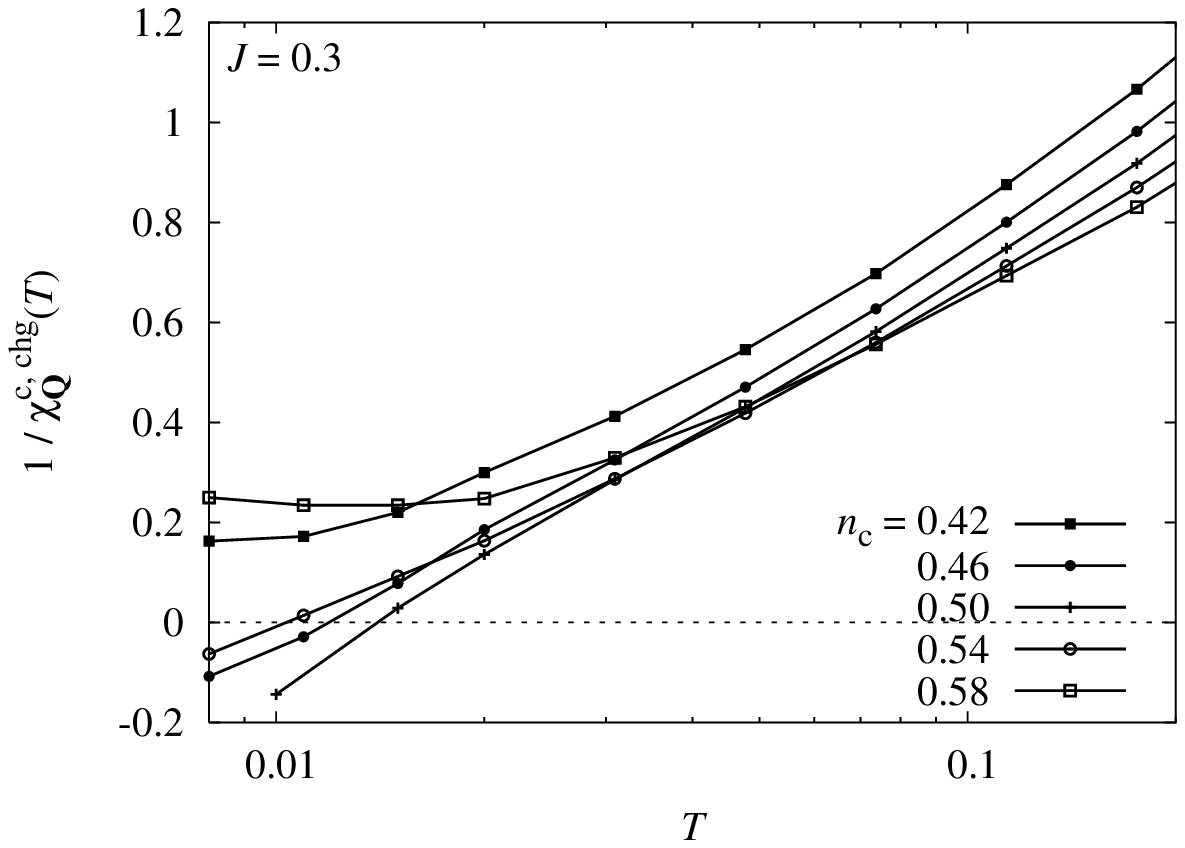}
	\end{center}
	\caption{
Temperature dependence of
inverse staggered charge susceptibility $1/\chi_{\mib{Q}}^{\rm c, chg}$ around the quarter filling for $J=0.3$.}
	\label{fig:KL-J030-T_suscep_c_stag}
\end{figure}

We next shift $n_{\rm c}$ from quarter filling. 
Figure~\ref{fig:KL-J030-T_suscep_c_stag} shows $1/\chi_{\mib{Q}}^{\rm c, chg}$ at $J=0.3$ for various 
fillings 
around $n_{\rm c}=0.5$. 
The transition temperature 
decreases on both sides off quarter filling. 
%
This result seems to be reasonable since the quarter filling is favorable for the CDW state at $\mib{q}=\mib{Q}$, where the conduction electrons tend to align at alternate sites. 
The mechanism of the CDW instability will be discussed in detail in \S\ref{sec:summary} based on the strong-coupling expansion.


\section{Ground-State Phase Diagram}

\begin{figure}[tbp]
	\begin{center}
	\includegraphics[width=\linewidth]{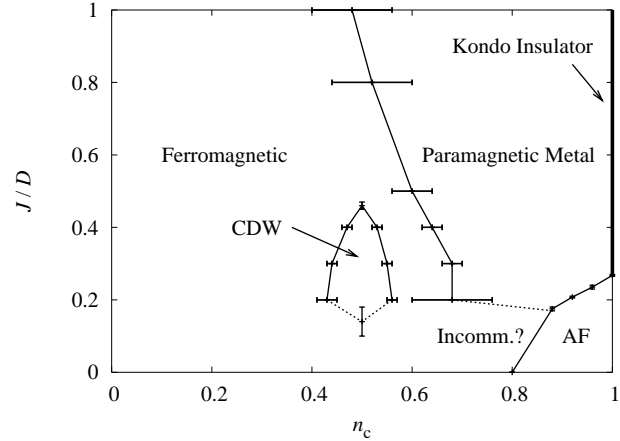}
	\end{center}
	\caption{
The ground-state phase diagram of the infinite-dimensional Kondo lattice. 
The dashed lines indicate possible boundaries. }
	\label{fig:KL-phase_diagram}
\end{figure}

Figure~\ref{fig:KL-phase_diagram} shows a ground-state phase diagram against $J/D$ and $n_{\rm c}$ estimated from the finite-temperature calculation. 
The boundaries are determined through scanning the inverse susceptibilities with varying either $J$ or $n_{\rm c}$, as in Figs.~\ref{fig:KL-T_suscep_stag_inv}, \ref{fig:KL-J040-T_suscep_unif_inv}, \ref{fig:KL-n025-T_suscep_c_stag} and \ref{fig:KL-J030-T_suscep_c_stag}. 
The error bars are parallel to the axis of the scanned parameter.

Near half filling, antiferromagnetic and paramagnetic phases divide into weak- and strong-coupling 
regimes. 
The boundary agrees 
roughly with the condition of comparable energy scales of the RKKY interaction and the Kondo effect, which demonstrates the validity of Doniach's picture. 
The paramagnetic state at half filling is the Kondo insulator, as revealed by the uniform charge susceptibility in Fig.~\ref{fig:KL-T_suscep1}.

The ferromagnetic phase 
appears for $n_{\rm c} \lesssim 0.6$,
which is naturally understood from the RKKY interaction. 
The polarization function takes a lager value at $\mib{q}=0$ 
rather than $\mib{q}=\mib{Q}$ as shown in Fig.~\ref{fig:suscep_free}. 
The ferromagnetic phase 
is enlarged as compared to the mean-field estimation of Doniach.
The ferromagnetic phase is stable beyond $J=1.3$ at $n_{\rm c}=0.2$. 

The CDW state at $\mib{q}=\mib{Q}$ arises around the quarter filling $n_{\rm c}=0.5$ with a range of about $\pm 0.06$.
We could not 
accurately 
determine the boundary on the weak-coupling side because of the exponentially decreasing 
transition temperature. Hence, possible boundary is represented by the dashed line in Fig.~\ref{fig:KL-phase_diagram}.  
Since the CDW state 
still keeps a part of the entropy of localized spins, 
the magnetic ordering is expected at temperatures lower than the CDW transition. 
The magnetic susceptibilities at temperatures higher than the CDW ordering show a tendency toward the ferromagnetism. 
In fact, the CDW phase is surrounded by the ferromagnetic phase. 
Therefore, it is 
likely that the CDW state coexists with the ferromagnetism in the ground state. 
For couplings smaller than $J \simeq 0.14$, 
the energy scale of the charge fluctuation is smaller than that of the ferromagnetism.
Hence, the CDW ordering is 
unlikely to occur in the 
small-$J$ case.

Around $n_{\rm c}=0.8$, no divergence of the susceptibilities has been observed for any $J$ down to 0.1.
In view of the fact that the Kondo temperature is exponentially small in the weak coupling regime,
it is highly probable that an instability actually takes places at certain ordering vector other than $\mib{q}=0$ and $\mib{Q}$.
The polarization function $\Pi_{\mib{q}}^0$ of free electrons in fact has a maximum between $\mib{q}=0$ and $\mib{Q}$ as shown in Fig.~\ref{fig:suscep_free}.
For $n_{\rm c}=0.6$ and 0.8, the peak is located close to $\mib{q}=\mib{Q}$. 
Hence, we reasonably expect the incommensurate order.
%
The ordering vector $\mib{q}$ can change successively, so that the 
boundaries will be complicated
between the ferromagnetic and incommensurate as well as between the incommensurate and antiferromagnetic phases.

\section{Summary and Discussion}
\label{sec:summary}

\subsubsection*{Insulating phases at half filling}
With the tight-binding band in the infinite-dimensional hypercubic lattice, 
divergence of the antiferromagnetic susceptibility at $\mib{Q}=(\pi, \cdots, \pi)$ has been demonstrated 
for a small-$J$ case.
With increasing coupling constant, the paramagnetic ground state takes over since antiferromagnetic fluctuations are suppressed by the Kondo effect.
The critical coupling constant agrees with 
that estimated by comparison between the Kondo temperature and the RKKY interaction. 
Hence, Doniach's picture is valid at half filling, although the estimation does not take account of formation of the Kondo-insulating gap.
We should note, however, 
that Doniach's picture fails in the regime far from half filling.

The antiferromagnetic state at half filling should be an insulator, since the Fermi surface with the nesting property disappears by the ordering. 
On the other hand, the 
paramagnetic regime at $J > J_{\rm c}$ has been shown to be the Kondo insulator. 
Consequently, the ground state at half filling is an insulator regardless of the coupling constant $J$. 
The mechanism of the energy gap, however, differs between 
large- and small-$J$ regimes.

\subsubsection*{Ferromagnetism at low carrier densities}
We have confirmed divergence of the ferromagnetic susceptibility in the low-carrier-density regime. 
The appearance of the ferromagnetism can be understood by the RKKY interaction. 
The variational method based on the mean-field theory has actually proposed its existence\cite{Lacroix-Cyrot, Fazekas-Muller-Hartmann}. 
On the other hand, 
in the infinite-dimensional Anderson lattice, 
Jarrell {\it et al.} did not find a divergent ferromagnetic susceptibility for any filling or set of parameters\cite{Jarrell-PAM}.
Hence, 
the localized spins seem to be required for ferromagnetism.

The ferromagnetic phase extends beyond the critical coupling constant estimated from the comparison between the Kondo temperature and the RKKY energy scale. 
This follows from the fact that all the local spins cannot be screened by a fewer 
number of conduction electrons\cite{nozieres}. 
The local susceptibility consequently follows the Curie-Weiss law even at the strong coupling.
Our results imply that the effective interaction between the local spins is ferromagnetic. 
Note that the above argument assumes the large number of interacting neighbors where the spin-pair correlation is less important.
In low dimensions, on the other hand, intersite correlations among localized spins can make a collective singlet even without screening of conduction electrons as in the one-dimensional antiferromagnetic Heisenberg model.

\subsubsection*{CDW state at quarter filling}
At quarter filling, we have found an instability of conduction electrons against CDW formation at $\mib{Q}$. 
The CDW ordering takes place in the middle range of the coupling constant. 
This 
work is the first confirmation of the CDW state in the Kondo lattice model to our knowledge. 

The origin of the CDW instability can be found in the strong-coupling effective model.
In the strong-coupling limit, each conduction electron forms the Kondo singlet to screen the local spin.
Since the spatial extent of the Kondo cloud is reduced to 
the lattice spacing at $J=\infty$, the Kondo singlet may be regarded as a hole of the local spin.
Accordingly, the strong-coupling limit of the Kondo lattice model can be mapped to the infinite-$U$ Hubbard model, where the number of itinerant electrons is $1-n_{\rm c}$ per site.
In the effective Hamiltonian of order $t^2/J$, we find a repulsion between the itinerant 
holes, or equivalently, between the Kondo singlets\cite{Hirsch, Sigrist}.
Hence, with decreasing $J$, the repulsion gets larger and may give rise to the CDW transition.
This mechanism was first proposed by Hirsch for one dimension\cite{Hirsch}.

Although the strong-coupling expansion predicts the CDW instability,
whether the CDW ordering actually takes place or not depends on the range where the strong-coupling picture 
remains valid.
Namely, if this picture breaks at $J$ higher than the critical value for the CDW transition, the CDW state 
may not appear. Our results show that the strong-coupling picture holds away from $J=\infty$ in infinite dimensions. 
The appearance of the CDW state can be intuitively understood 
as the gain of the Kondo energy by the alternate distribution of conduction electrons.


\subsubsection*{Remaining issues}
In the present calculation, we have discussed instabilities only at the vector $\mib{q}=0$ and $\mib{Q}$. 
As has already been noted, there is a possibility of the ordering at the vector other than $\mib{q}=0$ and $\mib{Q}$ for an intermediate filling. 
In the mean-field theory at weak coupling, the ordering is expected at the peak of the polarization function $\Pi_{\mib{q}}$, which can be incommensurate. 
It is an interesting issue whether the incommensurate ordering actually occurs or the commensurate vectors are chosen, when the local correlation is taken into account. 

The ground state in the CDW phase 
remains an open problem.
Although occurrence of the CDW ordering has been demonstrated, 
the CDW state does not fix the spin configuration nor screen all the localized moments.
In the strong-coupling picture, half the moments remain free at quarter filling. 
In order to address the magnetic correlation in the ordered state, we need to work with a two-sublattice model. 
The problem is whether the ferromagnetic correlation observed outside the CDW phase remains below the ordering temperature. 
The two-sublattice calculations also 
provide informations on electronic states in the antiferromagnetic phase
 as in the Hubbard model\cite{Georges-Krauth} and the Anderson lattice model\cite{Rosenberg}. 
Recently, the reduced antiferromagnetic moment in the Kondo lattice has been derived for the Bethe lattice\cite{Peters-Pruschke}.

Our formalism on the magnetic instabilities is based on the DMFT, so that the inter-site correlation is incorporated in the mean-field level, 
which is justified in infinite spatial dimensions.
In order to address two- or three-dimensional systems, spatial dependences of the self-energy as well as of the vertex part are indispensable. 
Extensions of the DMFT has been performed either by replacing the impurity with a cluster\cite{Maier}, or by reconstructing the self-energy from the local vertex evaluated in the effective impurity\cite{Kusunose, Toschi}.
Investigation of the two- or three-dimensional Kondo lattice model is 
a most challenging task, which should involve study of anisotropic superconductivity.

\section*{Acknowledgment}
We acknowledge useful discussions with N. Shibata and H. Yokoyama.
One of the authors (J.O.) was supported by Research Fellowships of the Japan Society for the Promotion of Science for Young Scientists.

\appendix

\section{The RKKY Interaction and Mean-Field Theory}
In this appendix, we derive the RKKY interaction, and estimate the 
magnetic transition temperature 
in the mean-field theory.
Using the second-order perturbation theory for the Hamiltonian~(\ref{eq:H_KL}),
we obtain the RKKY interaction of the form
\begin{align}
	H_{\rm RKKY} = - \sum_{\langle ij \rangle} J_{\rm RKKY}(i, j) \mib{S}_i \cdot \mib{S}_j,
\label{eq:H_RKKY}
\end{align}
where the coupling constant $J_{\rm RKKY}(i, j)$ is given by
\begin{align}
	J_{\rm RKKY}(i, j) = 2J^2 \Pi^0_{ij} = \frac{2 J^2}{N_0} \sum_{\mib{q}} \Pi^0_{\mib{q}} {\rm e}^{{\rm i} \mib{q} \cdot (\mib{R}_i - \mib{R}_j)}.
\end{align}
The factor 2 comes from the spin degree of freedom.

We analyze the RKKY model, eq.~(\ref{eq:H_RKKY}), in the mean-field theory.
The spatial correlations of localized spins, $\chi^{\rm (MF)}_{\mib{q}}$, are given by
\begin{align}
	\chi^{\rm (MF)}_{\mib{q}} = \frac{\chi^0_{\rm loc}}{1 - J_{\rm RKKY}(\mib{q}) \chi^0_{\rm loc}},
\end{align}
where $J_{\rm RKKY}(\mib{q}) = 2J^2 \Pi^0_{\mib{q}}$. 
Without the renormalization effect by the conduction electrons, the susceptibility of the local spin follows the Curie law, $\chi^0_{\rm loc}=1/4T$.
A divergence of $\chi^{\rm (MF)}_{\mib{q}}$ yields the instability of the paramagnetic states.
Consequently, the transition temperature $T_{\mib{q}}^{\rm (MF)}$ is determined by
\begin{align}
	T_{\mib{q}}^{\rm (MF)} = \frac{1}{2}J^2 \Pi^0_{\mib{q}}(T_{\mib{q}}^{\rm (MF)}).
\label{eq:Tc_mf}
\end{align}
We note that $\Pi_{\mib{q}}^0$ in the right-hand side of this equation depends on the temperature.

\end{document}